\documentclass[modern]{aastex61}
\usepackage{graphicx}
\usepackage{psfig}
\usepackage{natbib}
\usepackage{subfigure} 

\newcommand{\gsim}{\mbox{\hspace{.2em}\raisebox{.5ex}{$>$}\hspace{-.8em}\raisebox{-.5ex}{$\sim$}\hspace{.2em}}}
\newcommand{\lsim}{\mbox{\hspace{.2em}\raisebox{.5ex}{$<$}\hspace{-.8em}\raisebox{-.5ex}{$\sim$}\hspace{.2em}}}

\newcommand{\twCO}{$^{12}$CO}  \newcommand{\thCO}{$^{13}$CO}

\newcommand{\HII}{\mbox{H\,\textsc{ii}}}

      \newcommand{\ps}{\,{\rm s}^{-1}}
       
    \newcommand{\km}{\,{\rm km}}

\begin{document}

\title{
The Milky Way Imaging Scroll Painting (MWISP): Project Details
and Initial Results from the Galactic Longitude of $+$25\fdg8 to $+$49\fdg7
}

\shorttitle{MWISP CO survey between 25\fdg8 and 49\fdg7}

\correspondingauthor{Yang Su}
\email{yangsu@pmo.ac.cn}

\author[0000-0002-0197-470X]{Yang Su}
\affil{Purple Mountain Observatory and Key Laboratory of Radio Astronomy,
Chinese Academy of Sciences, Nanjing 210034, China}

\author{Ji Yang}
\affiliation{Purple Mountain Observatory and Key Laboratory of Radio Astronomy,
Chinese Academy of Sciences, Nanjing 210034, China}

\author{Shaobo Zhang}
\affiliation{Purple Mountain Observatory and Key Laboratory of Radio Astronomy,
Chinese Academy of Sciences, Nanjing 210034, China}

\author{Yan Gong}
\affiliation{Purple Mountain Observatory and Key Laboratory of Radio Astronomy,
Chinese Academy of Sciences, Nanjing 210034, China}
\affiliation{Max-Planck Institute f{\"u}r Radioastronomy, Auf dem H{\"u}gel 69,
53121 Bonn, Germany}

\author{Hongchi Wang}
\affiliation{Purple Mountain Observatory and Key Laboratory of Radio Astronomy,
Chinese Academy of Sciences, Nanjing 210034, China}

\author{Xin Zhou}
\affiliation{Purple Mountain Observatory and Key Laboratory of Radio Astronomy,
Chinese Academy of Sciences, Nanjing 210034, China}

\author{Min Wang}
\affiliation{Purple Mountain Observatory and Key Laboratory of Radio Astronomy,
Chinese Academy of Sciences, Nanjing 210034, China}

\author{Zhiwei Chen}
\affiliation{Purple Mountain Observatory and Key Laboratory of Radio Astronomy,
Chinese Academy of Sciences, Nanjing 210034, China}

\author{Yan Sun}
\affiliation{Purple Mountain Observatory and Key Laboratory of Radio Astronomy,
Chinese Academy of Sciences, Nanjing 210034, China}

\author{Xuepeng Chen}
\affiliation{Purple Mountain Observatory and Key Laboratory of Radio Astronomy,
Chinese Academy of Sciences, Nanjing 210034, China}

\author{Ye Xu}
\affiliation{Purple Mountain Observatory and Key Laboratory of Radio Astronomy,
Chinese Academy of Sciences, Nanjing 210034, China}

\author{Zhibo Jiang}
\affiliation{Purple Mountain Observatory and Key Laboratory of Radio Astronomy,
Chinese Academy of Sciences, Nanjing 210034, China}

\begin{abstract}
The Milky Way Imaging Scroll Painting (MWISP) project is an unbiased Galactic plane 
CO survey for mapping regions of $l=-10^{\circ}$ to $+250^{\circ}$ and $|b|\lsim$5\fdg2 
with the 13.7~m telescope of the Purple Mountain Observatory. 
The legacy survey aims to observe the \twCO, \thCO, and C$^{18}$O ($J$=1--0)
lines simultaneously with full-sampling
using the nine-beam Superconducting SpectroScopic Array Receiver
(SSAR) system with an instantaneous bandwidth of 1~GHz.
In this paper, the completed 250~deg$^2$ data from $l=+$25\fdg8 to $+$49\fdg7
are presented with a grid spacing of 30$''$ and a typical rms noise level of
$\sim$~0.5~K for \twCO\ at the channel width of 0.16~$\km\ps$ and 
$\sim$~0.3~K for \thCO\ and C$^{18}$O at 0.17~$\km\ps$.
The high-quality data with moderate resolution ($\sim$50$''$), uniform sensitivity,
and high spatial dynamic range, allow us to investigate the details of 
molecular clouds (MCs) traced by the three CO isotope lines. 
Three interesting examples are briefly investigated, including distant Galactic spiral arms 
traced by CO emission with $V_{\rm LSR} <$0~km~s$^{-1}$,
the bubble-like dense gas structure near the \HII\ region W40,
and the MCs distribution perpendicular to the Galactic plane.
\end{abstract}


\section{Introduction}
Molecular gas plays a crucial role in star formation.
Molecular hydrogen (H$_2$) is the dominant component of molecular clouds (MCs)
in the interstellar medium (ISM). 
Unfortunately, H$_2$ radiates inefficiently in the cold, dense molecular ISM due to
a lack of a permanent dipole moment and corresponding dipolar rotational transitions
in the radio band \citep[e.g.,][]{2013ARA&A..51..207B}.
Carbon monoxide (CO), which is the next most abundant molecule 
in the ISM, is widely used to trace the molecular gas because CO emission is 
easily excited in the molecular ISM environment and the CO $J$=1--0 transition 
at 2.6 mm (or 115~GHz) is readily observed on the ground 
(e.g., \citealp{1991ARA&A..29..195C,2001ApJ...547..792D,2015ARA&A..53..583H}
and references therein).

Systematic CO surveys of the Milky Way are helpful for
improving our knowledge of the molecular ISM, the physics of star
formation, and the Galactic structures.
So far, many Galactic CO surveys have been done using single-dish telescopes,
such as the FCRAO \twCO~($J$=1--0) survey of the Outer Galaxy \citep{1998ApJS..115..241H},
the CfA 1.2~m complete \twCO~($J$=1--0) survey 
\citep[see Figure~1 and Table~1 in][]{2001ApJ...547..792D},
the Bell Laboratories 7~m \thCO~($J$=1--0) survey \citep{2001ApJS..136..137L},
the NANTEN 4~m \twCO~($J$=1--0) survey \citep{2004ASPC..317...59M},
the FCRAO 14~m \thCO~($J$=1--0) Galactic Ring Survey \citep[GRS,][]{2006ApJS..163..145J},
the Mopra 22~m \twCO, \thCO, C$^{18}$O~($J$=1--0) survey \citep{2013PASA...30...44B},
and the Three-mm Ultimate Mopra Milky Way Survey (ThrUMMS) for 
\twCO, \thCO, C$^{18}$O, and CN~($J$=1--0) lines \citep{2015ApJ...812....6B}.

Detailed information for large CO surveys was
summarized in Figures~1--2 in \cite{2015ARA&A..53..583H}, in which
the authors enumerate the major surveys of \twCO\ and \thCO~($J$=1--0) 
emission along the Galactic plane from 1970--2015.
The FOREST Unbiased Galactic plane Imaging survey with
the Nobeyama 45~m telescope \citep[FUGIN;][]{2017PASJ...69...78U}
plans to conduct the simultaneous \twCO, \thCO, and C$^{18}$O~($J$=1--0) 
observations toward the Galactic plane.
This survey will achieve rather high angular resolution
($\sim 20''$) data of CO~($J$=1--0) lines for the Galactic plane.

The \twCO~($J$=3--2) High-Resolution Survey of 
the Galactic plane \citep[COHRS,][]{2013ApJS..209....8D} and the 
\thCO$/$C$^{18}$O~($J$=3--2) Heterodyne Inner Milky Way Plane Survey 
\citep[CHIMPS,][]{2016MNRAS.456.2885R} were made using the 15~m James 
Clerk Maxwell Telescope (JCMT) in the submillimeter wavelengths, with
an angular resolution of $\sim14''$ for COHRS and $\sim15''$ for CHIMPS,
respectively.
The SEDIGISM survey \citep[Structure, Excitation, and Dynamics of 
the Inner Galactic ISM,][]{2017A&A...601A.124S}
covers 78~deg$^2$ in the fourth quadrant using the \thCO\ and C$^{18}$O~($J$=2--1)
lines at the 1~mm band.
All of these CO spectral line surveys, together with other Galactic multiwavelength
legacy data, will lead to significant advances in our understanding 
of the ISM physics of our Galaxy.

The Milky Way Imaging Scroll Painting (MWISP\footnote{http://english.dlh.pmo.cas.cn/ic/}) 
project is an ongoing northern Galactic plane CO survey using the
13.7~m millimeter-wavelength telescope at Delingha, China (hereafter DLH telescope). 
The MWISP project is led by the Purple Mountain Observatory (PMO), with the 
full support from the staff members at Delingha.
The survey observes \twCO, \thCO, and C$^{18}$O~($J$=1--0)
lines simultaneously for regions of $l=-10^{\circ}$ to $+250^{\circ}$ and $|b|\lsim$5\fdg2
over 10 years.

In this paper, we present the results of the MWISP CO survey 
for regions of $l=+$25\fdg8 to $+$49\fdg7 and $|b|\lsim$5\fdg2.
Section 2 mainly describes the DLH telescope, the observing strategy
and data reduction, and the details of the MWISP project.
Section 3 displays the initial results for
the large-scale CO maps along the Galactic plane with a large
coverage in latitude.
Three interesting examples, including the most distant CO arm, the 
bubble-like C$^{18}$O structure near the \HII\ region W40, 
and the vertical distribution of CO gas,
are further investigated in that section.
Finally, Section 4 gives a summary and future prospects.

\section{The MWISP CO Survey}
\subsection{Telescope and Multibeam Receiver System}
The DLH telescope is
situated at approximately (37$^{\circ}$22\farcm4N, 97$^{\circ}$33\farcm6E),
close to Delingha, the third largest city of Qinghai Province in western China. 
At an altitude of about 3200~m,
dry and stable atmospheric conditions make  
Delingha an excellent site for millimeter astronomy.

The surface accuracy of the main reflector of the telescope is routinely measured 
and has a typical value of $\approx73$~$\mu$m.
The tracking accuracy of the telescope is
approximately 1$''$--3$''$ in both azimuth and elevation.
The pointing accuracy for the whole sky is better than 5$''$, 
about one-tenth of the telescope half-power beam width (HPBW) 
at 115~GHz ($\sim 50''$).
 
A 3$\times$3 multibeam sideband-separating Superconducting SpectroScopic Array Receiver
(SSAR) system \citep{Shan} was installed on the DLH telescope in 
the 2010 winter.
The centroids of the 3$\times$3 beam patterns have a regular spacing of 172$''$
\citep[see Figure~16 in][]{Shan}, which are also measured before every observational season.
The SSAR system employs two-sideband-separating superconductor-insulator-supercoductor 
mixers with a typical single sideband noise temperature of 60~K and 
image rejection ratio above 10~dB over the frequency range of 85--115~GHz.
The instrument includes Fast Fourier Transform Spectrometers (FFTSs), 
digital Local Oscillator (LO) sources, 
digital bias power supplies, and an independent intermediate frequency (IF) module.

The IF band is 2.64$\pm$0.5~GHz. The \twCO, \thCO, and C$^{18}$O ($J$=1--0) lines
at 115.271~GHz, 110.201~GHz, and 109.782~GHz, respectively,
are all within the 1~GHz band when 
the LO is set at 112.6~GHz.
Eighteen high-resolution FFTS digital spectrum analyzers
work at 1~GHz or 200~MHz bandwidth with 16,384 channels.
The 1~GHz bandwidth covers a wide velocity range for 
the CO line survey, while the 200~MHz bandwidth provides
higher spectral resolution observations for further study
\citep[e.g., see Table 1 in][]{2018A&A...620A..62G}.

Typical system temperatures, which include noises from the receiver, the antenna
(the optical system, the dome, and the membrane), and the atmosphere,
are $\sim$250~K for \twCO\ at the upper sideband and $\sim$140~K for 
\thCO\ and C$^{18}$O at the lower sideband, respectively.
Observations were calibrated using the standard chopping wheel method
that allows switching between the sky and an ambient temperature load.
Details on the telescope can be found at http://www.radioast.nsdc.cn/mwisp.php.

\subsection{Observations and Data Reduction}
The sky coverage of the MWISP project is divided into 10,941 cells for the region
of $l=-10^{\circ}$ to $+250^{\circ}$ and $|b|\lsim$5\fdg2.
Each cell with 30$'\times30'$ is scanned along the Galactic
longitude ($l$) and the Galactic latitude ($b$) at least twice
to reduce the fluctuation of noise.
We chose 30$'\times30'$ as the cell's size based on three factors.
One is that the cell can be completed in time
with the suitable scanning parameters. This consideration is important to 
ensure the atmospheric stability during the 1--1.5 hr observation for 
a full mapping (see below).
The other factor is to limit the final data size of the completed cell
for one line, e.g., $\sim $0.5~GB 
(91 pixels $\times$ 91 pixels $\times$ 16,384 channels $\times$ 4 Byte). 
Finally, such a size is good to divide the whole mapping sky 
with a regular form, leading to the easy labeling for a cell
in Galactic coordinates.

All observations are taken in position-switch On-The-Fly 
\citep[OTF; see][]{2018AcASn..59....3S} mode.
The scan rate is $50''$ (or $75''$) per second, with a dump time of
0.3 s (or 0.2 s), depending on a cell's elevations, off-positions,
and weather conditions.
The sampling interval is $15''$ (= scan rate $\times$ dump time)
and the spacing between scan rows is $10''$, fulfilling the oversampling
of the $50''$ beam of the 13.7~m telescope.
A cell thus can be fully mapped in one half an hour (or an hour)
along $l$ or $b$.
Generally, we used the (scan rate= $50''$~s$^{-1}$, dump time=0.3 s) mode in observations.
Mapping twice (or three hours) can well satisfy the expected rms noise level.
The (scan rate= $75''$~s$^{-1}$, dump time=0.2 s) mode is often used for some special regions
such as, for example, the low elevation cells toward $l<10^{\circ}-15^{\circ}$ 
(e.g., near the direction of the Galactic center),
or some cells with far off positions
(e.g., the Aquila Rift region and the Cygnus region).
Sometimes, scanning twice (e.g., three hours) cannot reach the 
expected rms noise level, so further mapping with the 
(scan rate= $75''$~s$^{-1}$, dump time=0.2 s) mode is also used to reduce the rms noise 
for the observed cells.

The most important consideration is to maintain uniform sensitivity,
which essentially depends on the weather or the total system temperature.
During observations, standard sources are thus observed about
every two hours in the position-switch mode.
The spectral profiles and intensities of standard sources are 
monitored daily to check the stability of the observations.

Before observations, the off position with an area of $\sim 8'\times8'$ 
(i.e., the sky coverage of the $3\times3$ beams) near the target 
cell (e.g., $\leq 2^{\circ}$) is carefully checked to ensure free of emission for 
the \twCO\ line.
However, the \twCO\ emission in the range
of 0--40~$\km\ps$ is widely distributed toward the inner Galaxy, especially 
for the central molecular zone, the Aquila Rift, and the Cygnus region.
When we observe these regions,
some off-positions with a little of \twCO\ emission ($T_{\rm MB}\lesssim$0.3~K) are
also used as the reference backgrounds.
For these regions, \twCO\ lines may show some weak absorption features at certain
local standard of rest (LSR) velocities. The relevant 
information was recorded for all observed cells.

The total bandwidth of 1~GHz, with 16,384 channels, provides 
a channel frequency interval of 61 kHz, resulting in a velocity
separation of about 0.16~km~s$^{-1}$ for \twCO\ and 0.17~km~s$^{-1}$
for \thCO\ and C$^{18}$O.
A first-order (or linear) baseline was fitted for the CO spectra.
Most of the bad channels in the spectra were removed.
The antenna temperature, $T_{\rm A}^{\star}$, 
was converted to the main beam brightness temperature, $T_{\rm MB}$,
with the formula of
$T_{\rm MB}=T_{\rm A}^{\star}/(f_{\rm b}\times\eta_{\rm MB})$.
The beam-filling factor of $f_{\rm b}$ is
assumed to be 1 for the extended CO emission and the main
beam efficiency $\eta_{\rm MB}$ varied between 40$\%$ and 50$\%$  
in the past seven seasons.
All intensities shown in this paper are on the $T_{\rm MB}$ scale.

Due to the field rotation, a larger cell size of 45\farcm5 $\times$ 45\farcm5 
was preserved in the data processing, resulting in some overlap between 
neighboring cells.
For each cell named as GLLL.L$\pm$BB.B, all CO spectra 
were summed together with the rms as the weight.
Finally, the three-dimensional (3D) FITS data cubes of each cell
were made with a grid spacing of $30''$ for \twCO, \thCO, and
C$^{18}$O ($J$=1--0) lines.
All data were reduced using the
GILDAS software\footnote{http://ascl.net/1305.010 or
http://www.iram.fr/IRAMFR/GILDAS} \citep{2005sf2a.conf..721P}.

\subsection{Characteristics and Advantages of the MWISP}
The MWISP project is a large-scale, unbiased, and high-sensitivity
triple CO isotope line survey for the northern Galactic plane
(e.g., $l=-10^{\circ}$ to $+250^{\circ}$ and $|b|\lsim$5\fdg2)
that uses the 13.7~m single-dish telescope.
The new nine-beam array with the OTF observation mode increases
the mapping speed by roughly an order of magnitude,
compared with the old single-beam system.

The DLH telescope can observe the northern Galactic plane for about
15--16 hr per day from September to April.
Generally, each cell with the required rms noise (see Table~1) consumes
approximately 3--5 hr (or 2--4 scans along $l$ and $b$).
That is, the MWISP project can map about 240--250 square degrees of the 
Galactic plane per year.
The MWISP survey started in 2011 November and is expected to be
completed in 2022.

The characteristics and advantages of the MWISP CO survey are summarized as follows:

1. Large-scale CO mapping with a high spatial dynamic range.
The MWISP project will provide us with large-scale CO maps of
$\sim2600$~deg$^2$, with a moderate angular resolution of
$\sim50''$. The final full-sampling 3D CO datasets have a grid spacing of $30''$.
Researchers thus can investigate the detailed structures of local MCs
(e.g., 1~pixel$\lsim$0.15~pc at a distance of $\lsim 1$~kpc),
the properties of the MCs with moderate distances (e.g., $\sim$3--6~kpc),
and the distribution of the distant molecular gas
(e.g., $\gsim$10~kpc).

2. Unbiased CO survey with high sensitivity. 
The unbiased survey with high sensitivity, wide velocity coverage,
and high velocity resolution has a uniform sensitivity, 
providing us a clearer picture of the molecular gas distribution and properties
of the Milky Way.
Through systematic studies of the CO emission of the molecular gas,
many new MCs will be revealed due to the high-quality data of the new survey,
which have an rms sensitivity of $\sim 0.5~(0.3)$~K,
a velocity coverage of $\sim 2600~\km\ps$, and a velocity resolution of 
$\sim 0.16~(0.17) \km\ps$ for \twCO~(\thCO\ and C$^{18}$O) lines.

3. Simultaneous observations of \twCO, \thCO, and C$^{18}$O~($J$=1--0) transitions. 
The \twCO\ emission tracing the 
total molecular gas can reveal structures and distributions of 
the diffuse gas with a typical density of $10^2$~cm$^{-3}$.
The optically thinner \thCO\ and C$^{18}$O lines trace denser 
molecular gas with a typical density of $10^3$--$10^4$~cm$^{-3}$
because of their less abundance with respect to H$_2$ and 
therefore less optical depth effects.
Therefore, the MWISP survey is adequate for revealing the diffuse gas of MC envelopes,
denser molecular gas of giant molecular clouds (GMCs), and the changes in
abundance ratio between \thCO\ and C$^{18}$O due to
isotope-selective photodestruction of the rarer CO species.

\begin{figure}[ht]
\includegraphics[trim=0mm 0mm 0mm 0mm,scale=1.3,angle=0]{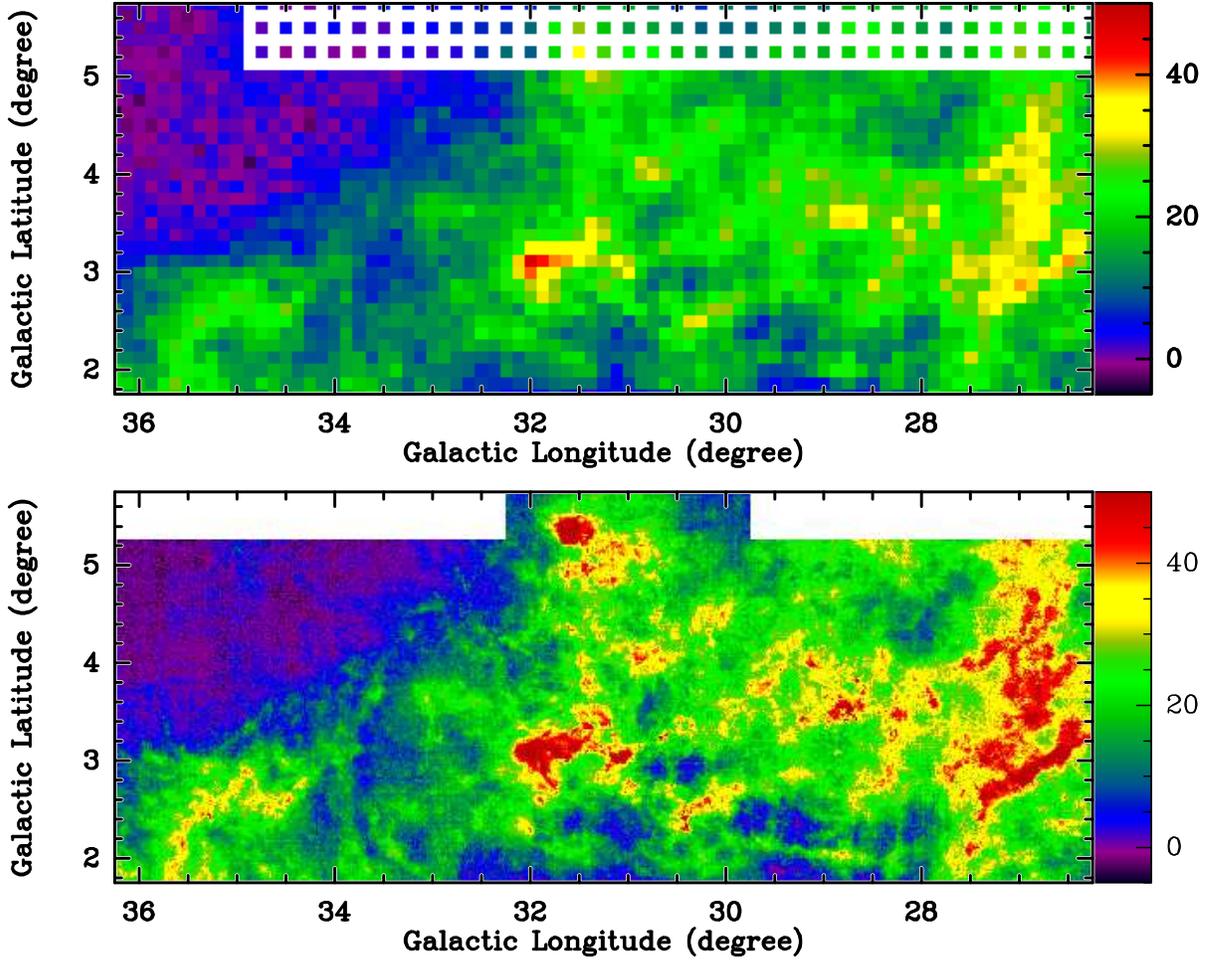}
\caption{
Top: The \twCO\ integrated intensity map of the Serpens$/$Aquila Rift MC complex 
from the Columbia-CfA CO survey \citep{2001ApJ...547..792D}.
Bottom: The MWISP \twCO\ integrated intensity map for the same region.
In both panels, all the integrated velocity ranges are 0 to 30~$\km\ps$,
and the color bars represent the integrated intensity in units of K~$\km\ps$.
\label{cfa-mwisp-aquila}}
\end{figure}

With a spatial resolution of $\sim 50''$ and a grid spacing 
of $30''$, the full-sampling MWISP survey provides a rich
CO data set for $b > 1^{\circ}$ regions, which are less covered by other CO
surveys, excluding the CfA 1.2~m complete \twCO~($J$=1--0) survey 
with a beam size of $\sim 8'$ \citep{2001ApJ...547..792D}.
Figure~\ref{cfa-mwisp-aquila} gives a comparison between the CfA 1.2~m CO map
and our MWISP map for the Serpens$/$Aquila Rift MC complex.
Note that the MWISP integrated map reveals more detailed structures than that of the
1.2~m CO data. Furthermore, the MWISP survey has a wider velocity coverage than 
that of the previous CO survey
\citep[e.g., $-5$ to $+$135~$\km\ps$ for the GRS \thCO\ survey,][]{2006ApJS..163..145J},
leading to more completed velocity coverage for MCs in the first quadrant of the 
Milky Way (e.g., the most distant MCs beyond the solar circle).

The ongoing FUGIN survey plans to investigate the distribution and 
properties of molecular gas in the Galaxy with the \twCO, 
\thCO, and C$^{18}$O ($J$=1--0) lines \citep{2017PASJ...69...78U}.
The region of $l=+10^{\circ}$ to $+50^{\circ}$ and $b=+1^{\circ}$ to $+1^{\circ}$
is fully covered using the multibeam (2$\times2$), dual-polarization, 
two-sideband receiver installed in the Nobeyama 45~m telescope.
Compared with the FUGIN project, the MWISP CO survey has the larger
coverage (2600~deg$^2$ versus 160~deg$^2$) and the higher velocity resolution
($0.16 \km\ps$ versus $0.65 \km\ps$). 
Meanwhile, the sensitivity of the MWISP survey is better 
than that of the FUGIN project \citep[e.g., $T_{\rm rms}(T_{\rm MB})\sim$~1.5~K 
for \twCO\ and $\sim$~0.7~K for \thCO$/$C$^{18}$O at 8\farcs5$\times$8\farcs5$\times1.3 \km\ps$ in the first quadrant regions,][]{2017PASJ...69...78U}. 
On the other hand, 
the spatial resolution of the Nobeyama 45~m telescope is about 2.5 times higher than 
that of the DLH 13.7~m telescope.
Therefore, the FUGIN data with the final grid spacing of 8\farcs5 can resolve more 
detailed MC structures in the Galactic plane of $|b| <1^{\circ}$.

In summary, the MWISP CO survey gives us
a good opportunity to study the Galactic structures,
the MC properties and the star formation, and the associations between
the molecular gas and the extended radio sources such as, for example, \HII\ regions
and supernova remnants (SNRs).
Much work has been done based on the MWISP data, such as Galactic
arms traced by the CO emission \citep{2015ApJ...798L..27S,2017ApJS..230...17S,
2016ApJS..224....7D,2017ApJS..229...24D,2016ApJ...828...59S},
MCs and star formation \citep{2013RAA....13..921L,2014AJ....147...46Z,
2015ApJ...811..134S,2016RAA....16...56Z,2016A&A...588A.104G,2017ApJ...835L..14G,
2016ApJ...822..114C,2017ApJ...838...80C,2017ApJ...838...49X,
2017ApJS..230....5W,2018ApJS..235...15L,2018ApJS..238...10L,Sun2019},
and interactions between SNRs and MCs
\citep{2014ApJ...788..122S,2014ApJ...796..122S,2017ApJ...845...48S,2017ApJ...836..211S,
2018ApJ...863..103S,2014ApJ...791..109Z,2016ApJ...833....4Z,2017A&A...604A..13C}.
The MWISP project also provides us an invaluable dataset for studying the
distribution and kinematics of the MCs in the Milky Way (Section 3).
The large-scale CO data are important in delineating the spiral structure
of the Milky Way \citep[e.g., see the recent review in][]{2018arXiv181008819X}.
More studies will be presented in the immediate future.

The MWISP survey is still ongoing. At the time of this writing the survey has completed 
about 60$\%$ of the planned 2600 square degree coverage 
(see the details at http://www.radioast.nsdc.cn/viewallobsed.php).
The complete set of the survey data generated by the MWISP project
will be made publicly available via an online archive. 
Early access to the 3D dataset for specific regions of the sky
is also possible through mutual collaboration.
People who are interested in the MWISP data can request the datacube by
contacting the email address dlhproposal@pmo.ac.cn. Many mosaic data cubes 
have already been provided to researchers through the FTP server.

\section{The Survey Data}
\begin{figure}
\gridline{
          \hspace{-8ex}
	  \fig{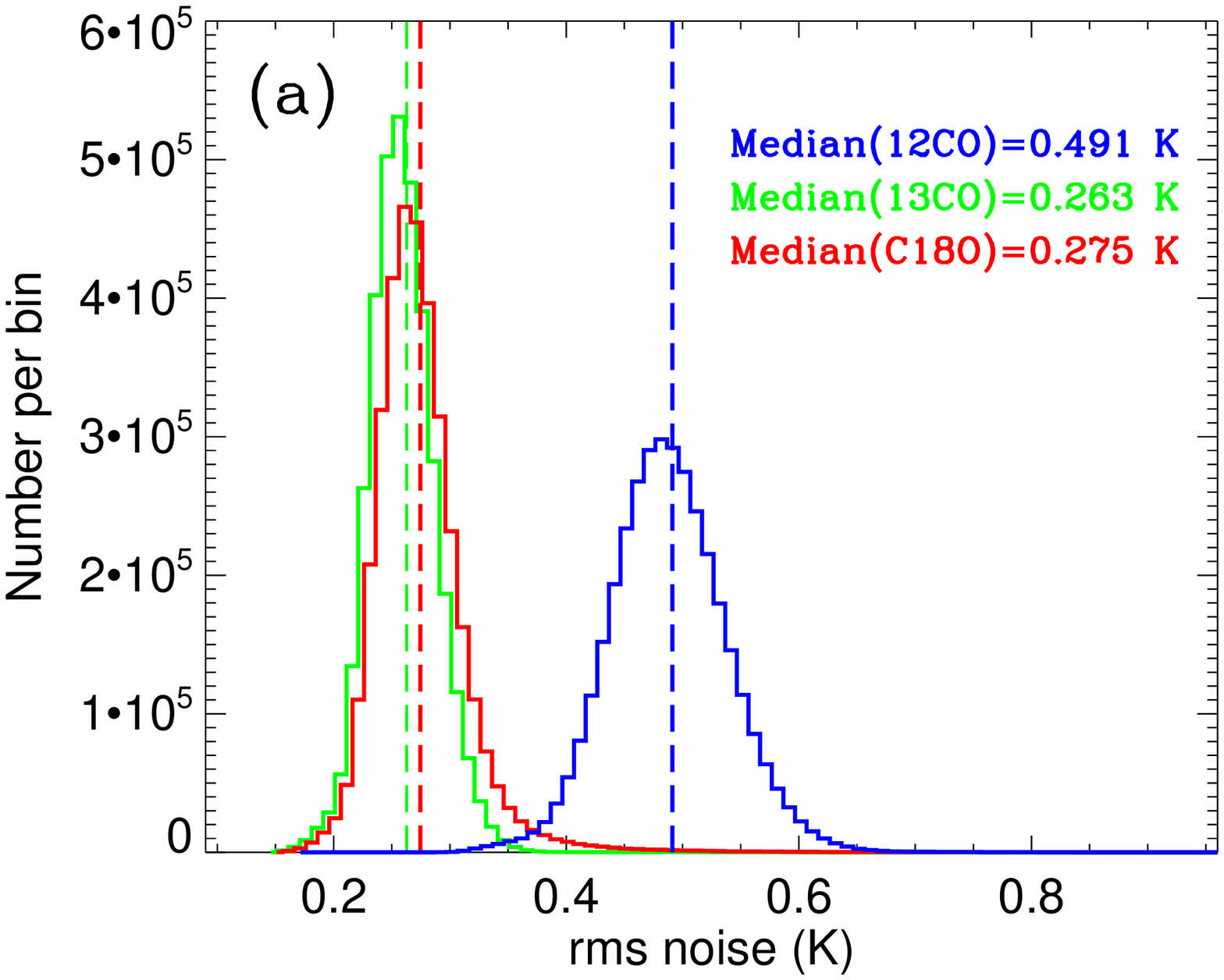}{0.6\textwidth}{}
	  \hspace{-5ex}
          \fig{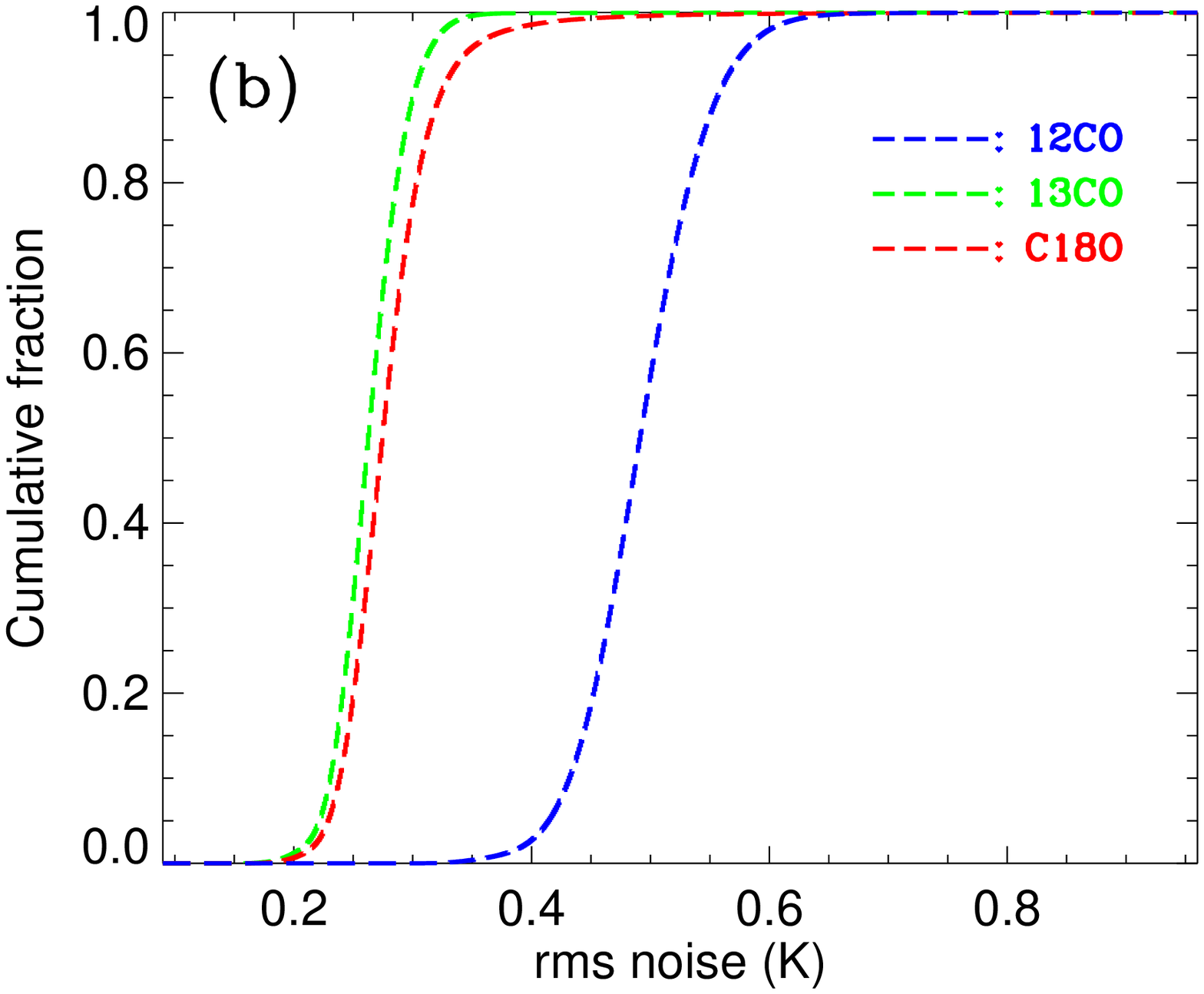}{0.6\textwidth}{}
          }
\caption{
(a): Distribution of rms noise values of the three lines for the whole dataset.
The long-dashed lines indicate the median values of \twCO~($J$=1--0) (blue),
\thCO\ ($J$=1--0) (green), and C$^{18}$O (red) emission, respectively.
(b): Cumulative distribution of rms noise values for the three CO lines.
\label{f1}}
\end{figure}

\begin{figure}
\gridline{
\hspace{1ex}
\fig{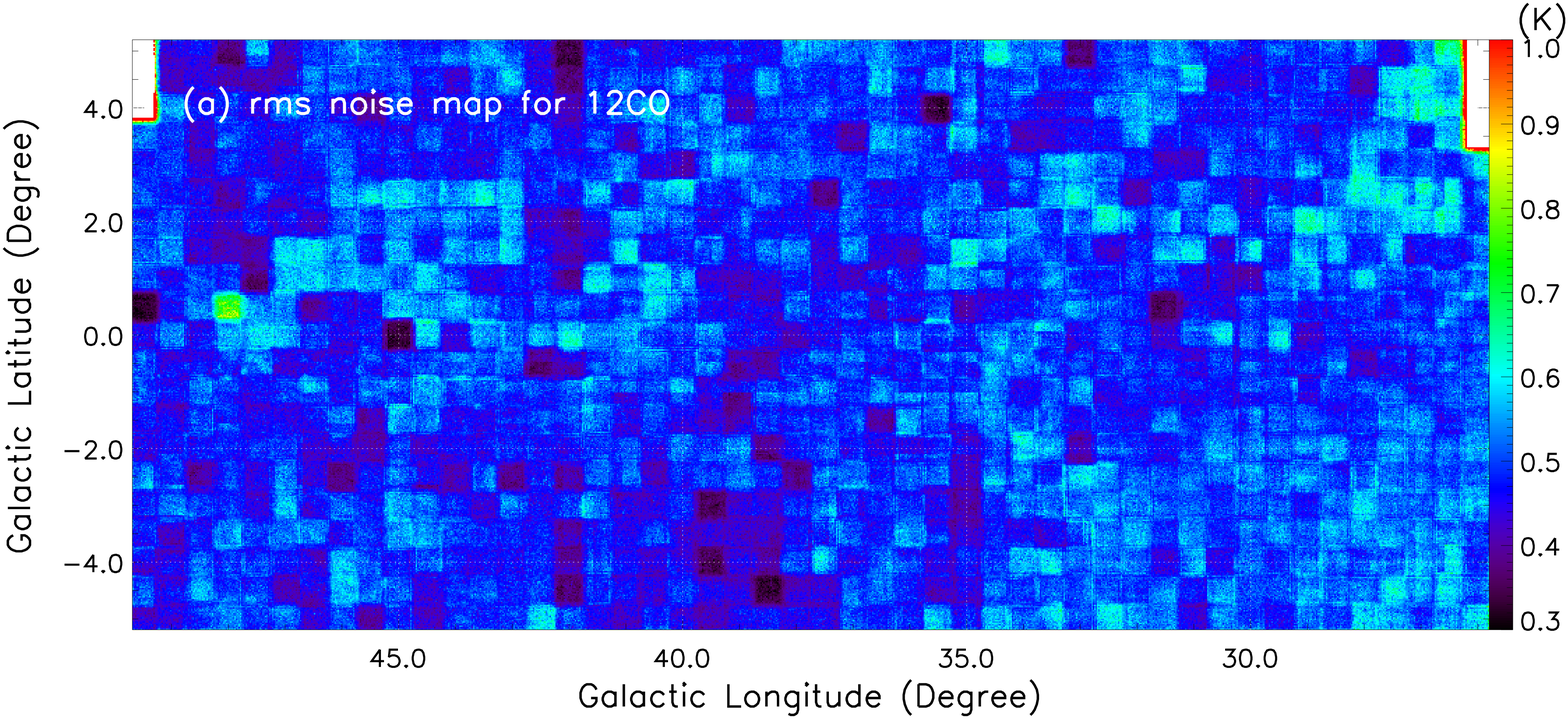}{1.\textwidth}{}
          }
\vspace{-8ex}
\gridline{
\hspace{1ex}
\fig{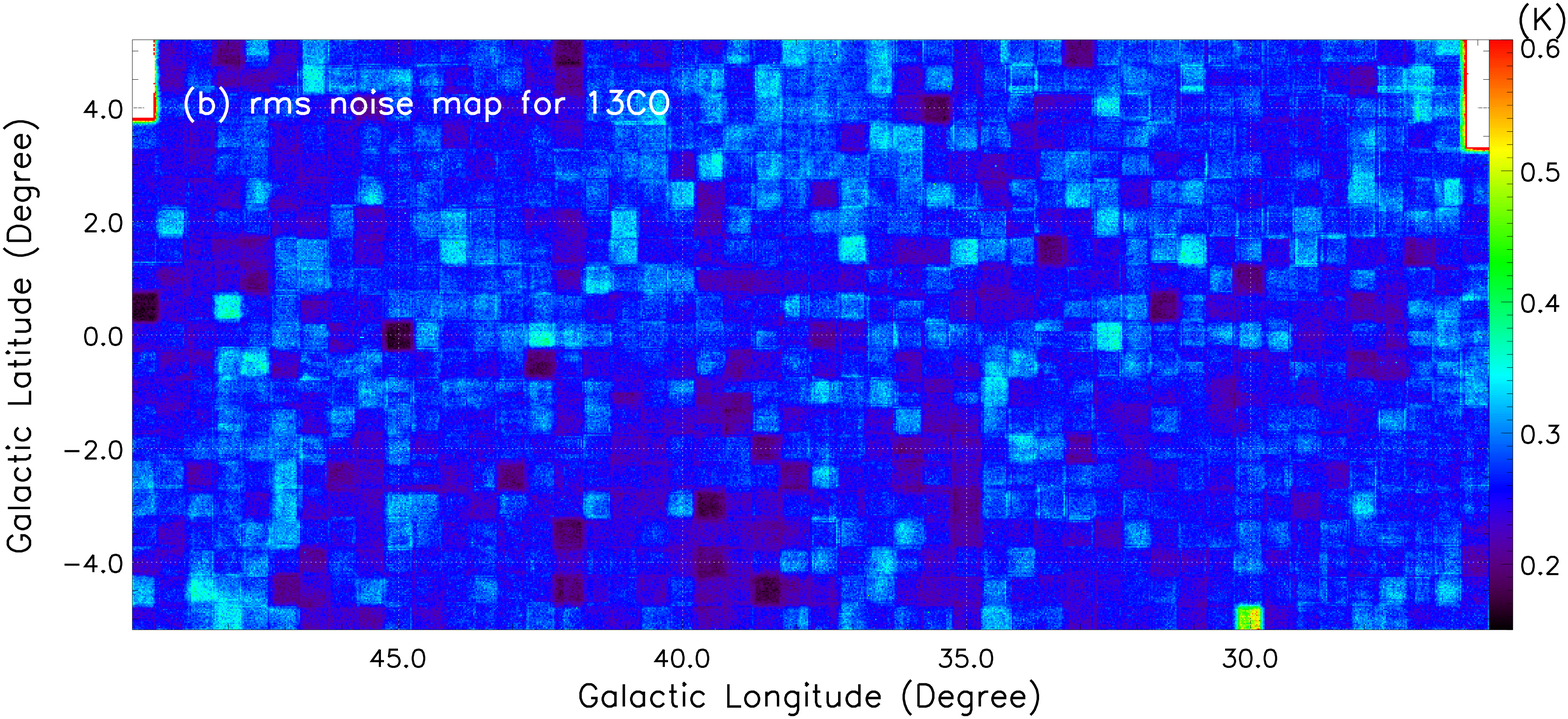}{1.\textwidth}{}
          }
\vspace{-8ex}
\gridline{
\hspace{1ex}
\fig{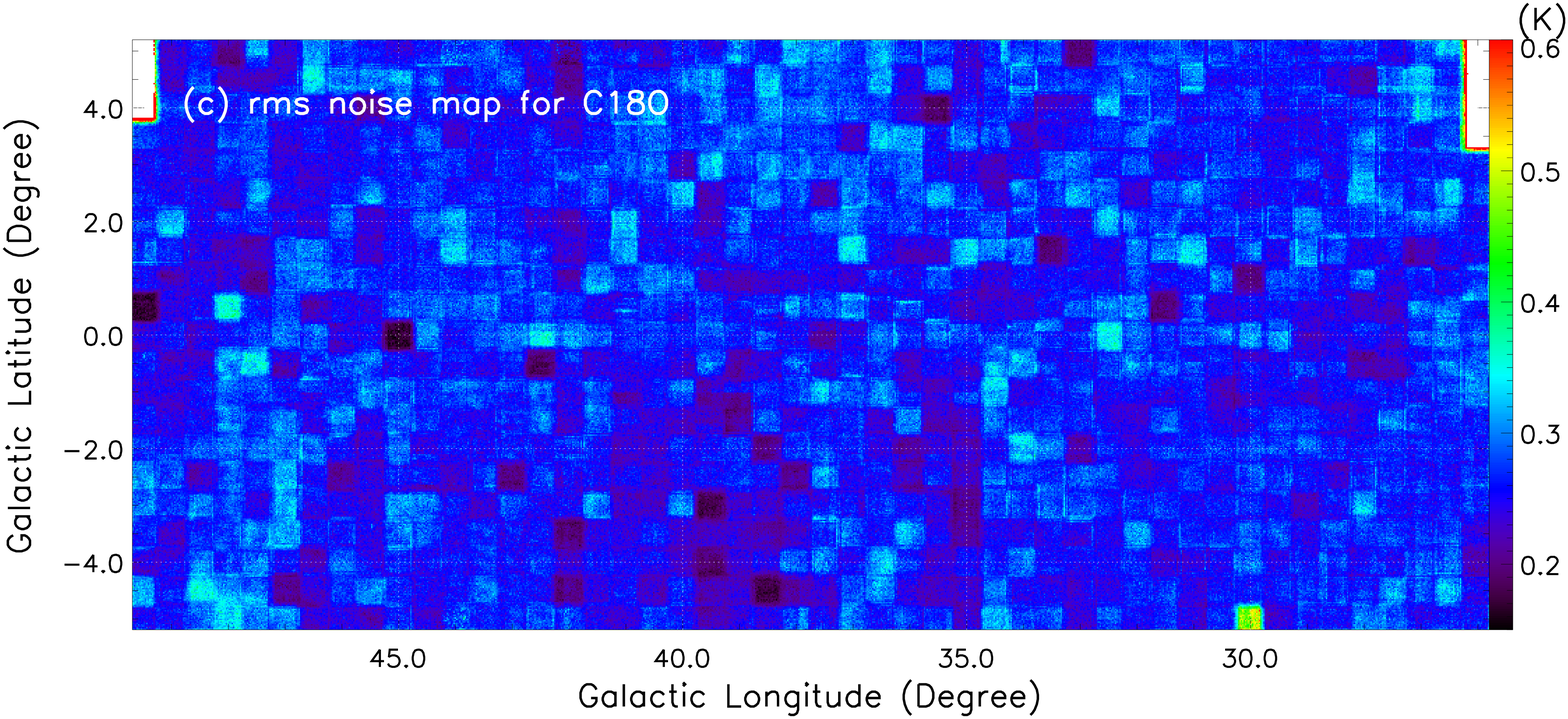}{1.\textwidth}{}
          }
\caption{
Rms noise maps for the \twCO, \thCO, and C$^{18}$O ($J$=1--0) lines. 
The maps display the variation in rms between cells. 
There are some stripes along $l$ and $b$ in the maps due to the 
bad weather (or the high system temperature) in the OTF mapping.
\label{rms}}
\end{figure}

In this section, we present the CO data for
the completed region of $l=+$25\fdg8 to $+$49\fdg7 and $|b|\lsim$5\fdg2.
Figure~\ref{f1} shows the rms distribution of the
$\sim$250 square degree dataset.
The typical rms noise level of the spectra is $\sim$0.5 K for
\twCO\ ($J$=1--0) at a channel width of 0.16~$\km\ps$ and $\sim$~0.3~K
for \thCO\ ($J$=1--0) and C$^{18}$O ($J$=1--0) at 0.17~$\km\ps$,
with a spatial resolution of $\sim50''$.
Generally, the rms noise of each cell is uniform for the 
three CO isotope lines in the 30$'\times30'$ region. 
As shown in Figure~\ref{rms}, however, the variation in rms
indeed exists between cells, e.g., $\sim $0.3--0.6~K for \twCO\ and
$\sim $0.15--0.35~K for \thCO/C$^{18}$O. Some of bright stripes with somewhat
larger rms noises (e.g., $\sim $0.6--0.8~K for \twCO\ and
$\sim $0.4--0.6~K for \thCO/C$^{18}$O) also can be seen 
along $l$ and $b$, with lengths of several tens of arcminutes.
These features are due to the bad weather (or the large system temperature) 
in the OTF scanning.
Table~1 lists the parameters of the CO data used in this paper.

The distributions of the CO emission are shown through intensity maps, 
intensity-weighted mean velocity maps, and position$-$velocity (PV) diagrams.
Some interesting results of the new CO survey
are also investigated. 
The CO data are smoothed to the 0.5~km~s$^{-1}$ velocity resolution
in order to improve the sensitivity for weak emission.
The improved rms levels of the analyzed data are
of $\sim$0.28~K for the \twCO\ line 
and $\sim$0.16~K for the \thCO\ and C$^{18}$O lines, respectively.

\subsection{CO gas with $V_{\rm LSR} <$0~km~s$^{-1}$}
Very recently, \cite{2016ApJ...828...59S} and \cite{2017ApJS..230...17S} 
presented the results of distant MCs 
traced by CO emission between $l=$34\fdg75 and $l=$45\fdg25.
These MCs with negative velocities are divided into 
the distant Outer Arm and the Extreme Outer Galaxy (EOG), respectively.
These two parts of the molecular gas are very likely the extension of  
the Norma--Cygnus Arm and the Scutum--Centaurus Arm in the 
first quadrant (i.e., molecular gas structures from the fourth quadrant of the
inner Galaxy to the first quadrant of the outer Galaxy).
Actually, the more negative velocity feature at $V_{\rm LSR} <$0~km~s$^{-1}$ 
was proposed to be from the Outer Scutum-Centaurus Arm 
\citep{2011ApJ...734L..24D}.

\begin{figure}
\gridline{
\hspace{-5ex}
\fig{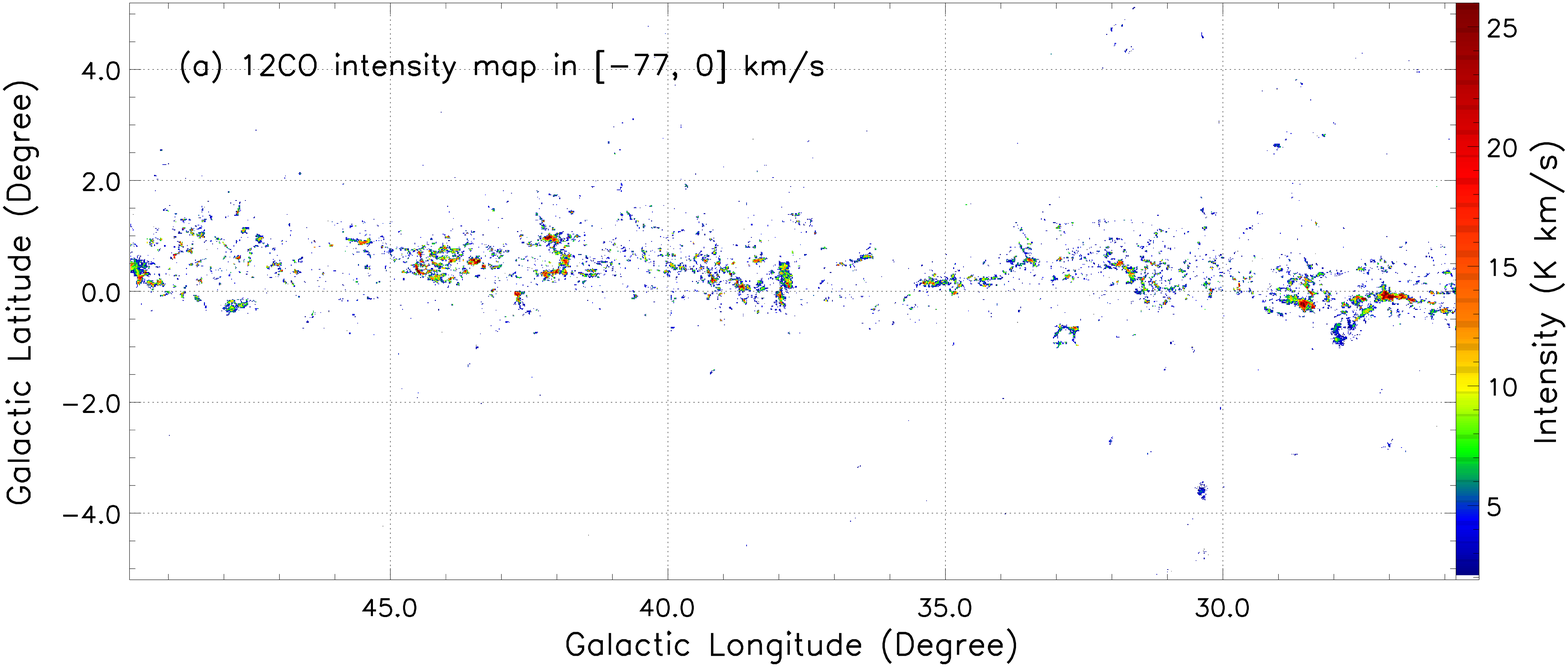}{1.2\textwidth}{}
          }
\vspace{-8ex}
\gridline{
\hspace{-5ex}
\fig{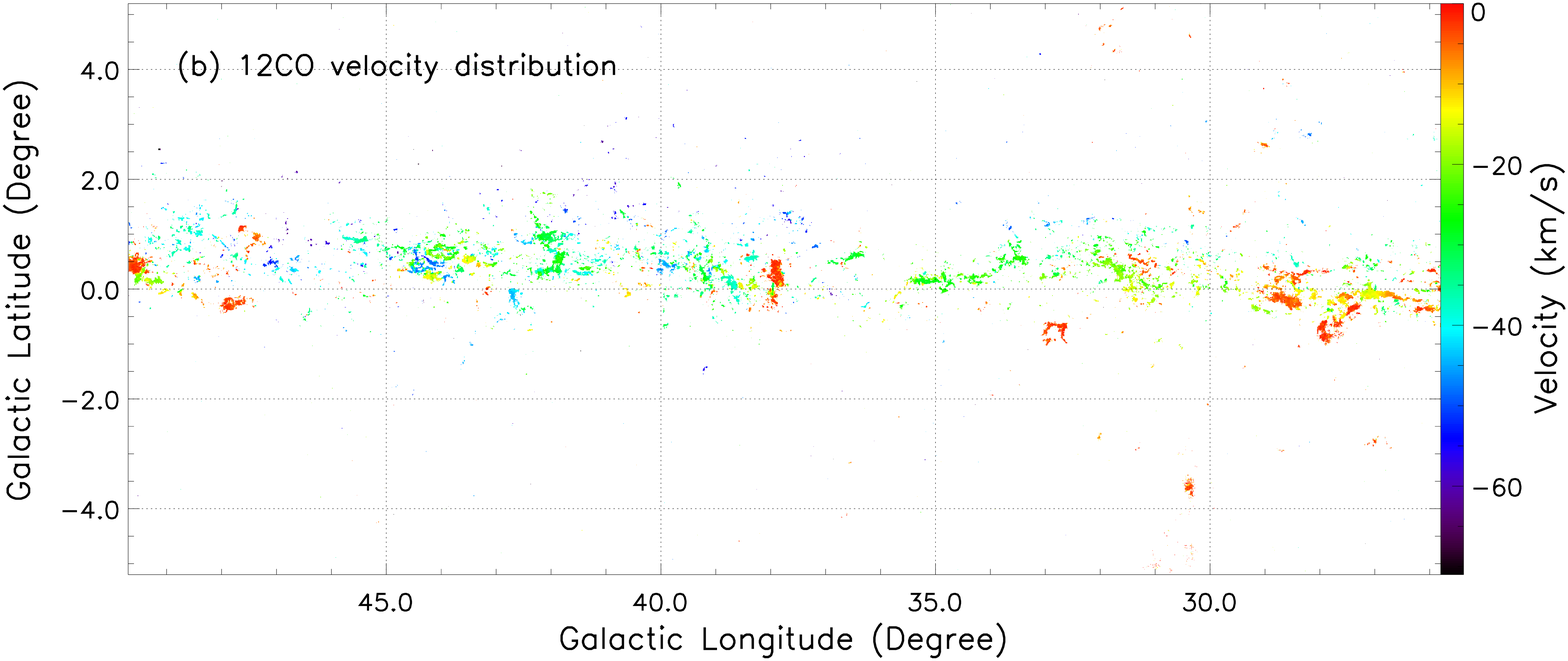}{1.2\textwidth}{}
          }
\caption{
(a) Integrated emission of the \twCO\ ($J$=1--0) emission
in the interval of $-77$ to 0~km~s$^{-1}$.
(b) Intensity-weighted mean velocity (first moment) map of the
\twCO\ emission for $V_{\rm LSR}\leq$0~km~s$^{-1}$ MCs.
\label{f2}}
\end{figure}

With the progress of the MWISP project, larger mapping was completed 
for the range of $+$25\fdg8$\lsim l \lsim +$49\fdg7 and $|b|\lsim$5\fdg2.
Figure~\ref{f2} shows the
spatial distribution of the CO gas with $V_{\rm LSR} <$0~km~s$^{-1}$,
together with the corresponding LSR velocity information.
A large amount of new \twCO\ emission is
revealed because of the high sensitivity and the large coverage of the
unbiased survey. 
Some of the MCs with relatively strong \twCO\ emission
have the corresponding \thCO\ emission, but none of them show significant
C$^{18}$O emission under the improved rms level of $\sim$~0.16~K
at the smoothed velocity resolution of 0.5~km~s$^{-1}$.
Higher resolution and sensitivity observations are expected to detect the 
weak emission of the dense gas far away from us.

Some of MCs with $V_{\rm LSR} \lsim$~0~km~s$^{-1}$ are likely the local molecular 
gas due to their $\sim$~0~km~s$^{-1}$ LSR velocities and relatively high 
$|b|$ values (e.g., $|b| \gsim$~2\fdg5; see the Aquila Rift region 
near $l\sim 26^{\circ}-32^{\circ}$ in the lower panel of Figure~\ref{f2}).
However, most of $V_{\rm LSR} <$~0~km~s$^{-1}$ MCs in Figure~\ref{f2} are 
believed to lie beyond the solar circle.
Obviously, CO emission is mainly concentrated near the Galactic plane,
excluding the possible local gas in the Aquila Rift region at high $|b|$.
In spite of this, the distant CO gas seems to also be slightly displaced from $b= 0^{\circ}$, 
which was explained by the warped plane at larger Galactocentric distances
and the apparent tilted structure caused by the Sun's $z$-height above the 
physical midplane of the
Galactic disk \citep[see][]{2016ApJ...828...59S,2017ApJS..230...17S}.

Due to the large distances, many MCs 
have small angular sizes and weak CO emission. These MCs are relatively 
isolated in $l$-$b$-$v$ space. 
Despite these properties, some interesting distant MCs with brighter CO 
emission also display 
extended concentrations, which usually have special morphologies 
such as filaments, arcs/shells, and irregular cavity-like structures.
These extended concentrations could be related to the massive star-forming 
activities in the surroundings.
Very recently, \cite{2018ApJ...852....2W} investigated the CO properties toward star-forming 
regions in the the Outer Scutum-Centaurus Arm using single pointing observation
of the 12~m telescope located at Arizona Radio Observatory.
Benefiting from the full CO map and the high sensitivity of the MWISP data,
we will systematically study the relationship between the distant MCs 
and the surrounding star formation activities.
Detailed identification and analyses of these MCs will be presented in future 
papers. Section 3.3.1 shows some initial results for the CO gas at
the edge of the Galaxy.

\subsection{CO gas with $V_{\rm LSR} >$0~km~s$^{-1}$}
Figure~\ref{guidemap} shows the \thCO~($J$=1--0) intensity map in the 0--130~km~s$^{-1}$
interval, overlaid with radio continuum contours from the Effelsberg
11~cm survey \citep{1990A&AS...85..633R}.
Some bright and/or extended radio sources discussed below 
are labeled on the guide map.

Figures~\ref{f3a} and \ref{f4a} show \twCO\ and \thCO\ 
channel maps, respectively. The channel maps are made
by integrating emission over a 10~km~s$^{-1}$ velocity bin
from 0~km~s$^{-1}$ to 120~km~s$^{-1}$. The 120--130~km~s$^{-1}$
\twCO\ and \thCO\ gas, which displays emission only 
near $l \sim 26^{\circ}-31^{\circ}$, is presented in Figure~\ref{f5}.
A large number of MC structures and features are seen in both of
the \twCO\ and \thCO\ channel maps. 

For the 0--20~km~s$^{-1}$ maps, the most prominent features
are the Aquila Rift \citep[e.g., see Figure~3 in][]{1985ApJ...297..751D},
which displays large-scale, diffuse, and enhanced CO emission 
in the field of view (FOV).
The MWISP survey, with the wide spatial dynamic range, reveals
lots of detailed structures for the large-scale molecular gas not
far from us \citep[i.e., $d \lsim$~450~pc;][]{2017ApJ...834..143O}.
Large-scale extended structures are also discernible from the 
PV diagrams of \twCO\ and \thCO\ emission (see the region of 
$V_{\rm LSR}\sim$~0--20~km~s$^{-1}$ and $l\sim 26^{\circ}$--$40^{\circ}$ 
in Figure~\ref{f6}). 
The \twCO\ $J$=2--1 and \thCO\ $J$=2--1 emission from the Aquila Rift region 
was investigated by \cite{2017ApJ...837..154N} using the 1.85~m 
telescope. Combinations of our CO $J$=1--0 data and the 1.85~m CO $J$=2--1
data will help us understand the detailed molecular gas properties of such
local regions.

Many of bright CO concentrations in the velocity interval of $\sim$~0--30~km~s$^{-1}$  
are found to be near the Galactic plane of $b \sim -$0\fdg4 to $+$0\fdg6, 
excluding the diffuse and extended emission from the local MCs
(e.g., the Aquila Rift region discussed above).
These concentrations, which have typical angular sizes of several arcminutes,
are likely related to the distant Perseus Arm in the first quadrant of the Galaxy.

For 20--40~km~s$^{-1}$ gas, the \twCO\ and \thCO\ emission
is found to be extended over a large region from $l\sim 34^{\circ} -44^{\circ}$
and $|b|\lsim 5^{\circ}$ in the channel maps.
This GMC complex is less studied in the literature. 
A 170~pc long giant molecular filament (GMF) of
G40.82$-$1.41, which is at $\sim$2~kpc 
\citep[e.g., the kinematic distance and the extinction distance,][]{2018ApJ...863..103S},
is probably a part of the GMC complex. The extended molecular gas
at $b \gsim 0^{\circ}$ of the GMC is related to 
the \mbox{H\,\textsc{ii}} regions Sh~2-75 (at $l=$40\fdg12, $b=$1\fdg50) 
and Sh~2-76 (at $l=$40\fdg44, $b=$2\fdg45).
One of the \mbox{H\,\textsc{ii}} regions, Sh~2-76, is exactly located at a distance of 
1.92$^{+0.09}_{-0.08}$~kpc from the parallax measurements of 
0.521$\pm 0.024$~mas \citep{2017MNRAS.466.4530C}.

At 30--60~km~s$^{-1}$, significant 
CO emission appears at $l\sim 34^{\circ}$--$36^{\circ}$ and 
$b\sim -3^{\circ}$ to $+1^{\circ}$, which is roughly perpendicular
to the Galactic plane.
Enhanced CO emission in the GMC complex displays complicated 
morphologies, such as multi-shells, bubbles, and cometary bright-rimmed 
structures. The GMC complex is associated 
with the high-mass star-forming region W48 (G35.2$-$1.8) at a parallax distance of 
3.27$^{+0.56}_{-0.42}$~kpc
\citep{2009ApJ...693..419Z}. 
The very bright SNR G34.7$-$0.4, also known as W44 at a near kinematic distance 
of $\sim$3.3~kpc \citep[e.g.,][]{2014IAUS..296..372S}, may be associated with 
the GMC complex.

For the $V_{\rm LSR}\gsim$50~km~s$^{-1}$ molecular gas, 
CO emission is mainly confined within $|b|\lsim 1^{\circ}$.
In addition to the CO emission close to the Galactic plane,
considerable MCs are detected in
$1^{\circ} \lsim |b| \lsim 2^{\circ}$, even for velocities near the tangent point
(see 80--120~km~s$^{-1}$ maps in Figures~\ref{f3a} and \ref{f4a}).
These features have not been revealed by previous CO surveys 
\citep[e.g., FCRAO GRS project,][]{2006ApJS..163..145J}
due to the limited Galactic latitude coverage of $|b| \lsim 1^{\circ}$
toward the first quadrant of the Milky Way.
From Figure~\ref{f6}, we find that the PV diagram of the MWISP \thCO\ data
is consistent with the previous 
PV map of the GRS \thCO\ survey \citep[see Figure~3 in][]{2006ApJS..163..145J}.
Furthermore, some new features in the \thCO\ PV diagram are 
unveiled for the $V_{\rm LSR}\lsim$50~km~s$^{-1}$ range 
(e.g., see structures at $l \sim 38^{\circ}$--42$^{\circ}$ and $V_{\rm LSR}\sim$25--45~km~s$^{-1}$,
the rectangle in the diagram)
because of the larger latitude coverage of the MWISP survey.
The CO emission of the structure is from the molecular gas
associated with \mbox{H\,\textsc{ii}} regions Sh~2-75 
and Sh~2-76.

Figure~\ref{f7} displays the spatial and velocity distribution of the
C$^{18}$O ($J$=1--0) emission in the velocity range of 0--40~km~s$^{-1}$.
Enhanced C$^{18}$O emission is clearly seen near $l\sim 28$\fdg7 and $b\sim 3$\fdg7,
which is probably associated with the feedback of massive stars in the Serpens south and 
the W40 complex (see Section 3.3.2).
Excluding C$^{18}$O emission from the 
local MCs such as the Aquila Rift, the rest of C$^{18}$O gas is mainly 
confined in $b\sim -2^{\circ}$ to $b\sim 3^{\circ}$.
In the figure, two other prominent C$^{18}$O features, which are located at 
$l \sim 35^{\circ}$ and $l \sim 40^{\circ}$, are found to be away from the 
$b =0^{\circ}$ plane. The two C$^{18}$O features are related
to the massive star-forming region W48 and \mbox{H\,\textsc{ii}} regions Sh~2-75 
and Sh~2-76, respectively.

Figure~\ref{f8} shows the C$^{18}$O emission in the velocity interval
of 40--80~km~s$^{-1}$ and 80--120~km~s$^{-1}$,
which is mainly within $|b| \lsim 2^{\circ}$ and $|b| \lsim 1^{\circ}$, respectively.
For the 40--80~km~s$^{-1}$ maps, the W48 GMC complex at $l \sim35^{\circ}$, 
which is roughly perpendicular to the $b =0^{\circ}$ plane,
is clearly seen in Figures~\ref{f8}a and \ref{f8}b.
On the other hand, the W51 GMC complex at a distance of 5.41$^{+0.31}_{-0.28}$~kpc
\citep[e.g., the trigonometric parallax from][]{2010ApJ...720.1055S}
also can be seen at $l \sim49^{\circ}$.
The  massive star-forming region W51 was recently reviewed by 
\citet[and references therein]{2017arXiv170206627G}.

For the $V_{\rm LSR}=$80--120~km~s$^{-1}$ maps (see Figures~\ref{f8}c,d), 
the most enhanced C$^{18}$O emission with $\sim 10$~K~km~s$^{-1}$
is associated with the mini-starburst region W43 at ($l\sim$30\fdg7, $b\sim$0\fdg0). 
The massive star-forming region, which is 
at a parallax distance of 5.49$^{+0.39}_{-0.34}$~kpc \citep{2014ApJ...781...89Z},
is probably at or close to the near end of the Galactic long bar.
The ongoing star-forming activity of the global mini-starburst region 
is likely the result of the massive gas clouds accumulated by the bar potentials 
and kinematics.
The gas property of the interesting GMC complex W43 was studied 
by many groups \citep[e.g.,][]
{2011A&A...529A..41N,2013A&A...560A..24C,2014A&A...571A..32M,2018PASJ..tmp..102S}.

For the survey data presented here, the C$^{18}$O emission is relatively
discrete in comparison with the extended \twCO\ and \thCO\ emission.
Researchers thus can identify the main structures of MCs using C$^{18}$O emission
due to the less velocity crowding and line blending
(e.g., see the W40 region in Section 3.3.2).

As shown in Figure~\ref{f8}, the distribution of the C$^{18}$O emission 
is below the plane of $b=0^{\circ}$, especially for the 40--80~km~s$^{-1}$ MCs.
Near the tangent point, where the dense gas is more concentrated in the plane,
the dominant part of them is also below the plane
(e.g., the 80--120~km~s$^{-1}$ maps).
This interesting feature is explained by the 
Sun's offset above the physical midplane of the Milky Way
\citep[e.g., $z_{\rm Sun} \sim$17.1~pc; see discussions and Figure~7 in]
[]{2016ApJ...828...59S}.
We further investigate the phenomena using MC samples 
close to the tangent point, e.g., the CO gas with $V_{\rm LSR} >$60~km~s$^{-1}$
(Section 3.3.3).

\subsection{Interesting Examples}
\subsubsection{CO gas at the Edge of the Milky Way}
The CO emission of the distant molecular gas is very weak.
Due to the warping and flaring of the
outer gas disk, the distant MCs with negative velocities are concentrated at the somewhat
higher latitudes, which is different from the molecular gas
in the first quadrant of the inner Galaxy. 
The unbiased MWISP CO survey has a wide velocity and areal coverage,
meanwhile there is also a high sensitivity, allowing us to 
systematically investigate weak CO emission far away from the Sun 
(see Figure~\ref{f2}).
Compared with the CfA 1.2~m CO survey \citep{2001ApJ...547..792D},
the MWISP CO data from the 13.7~m telescope obviously reveal detailed distributions
and structures of the distant MCs.

Figure~\ref{f9} shows the longitude--velocity ($l$--$v$) diagram
of the \twCO\ emission for the $V_{\rm LSR}\lsim$0~km~s$^{-1}$ molecular gas.
Note that the CO intensity is multiplied by a factor of 100 for the 
corresponding signal velocity range but not the whole velocity range.
The noise is therefore suppressed and the tiny features of weak CO emission 
are enhanced in the $l$--$v$ diagram. Based on the diagram,
we find that the bright emission  (thick blue parts in the figure) 
exhibits a large-scale molecular gas structure outside the solar circle.
On a large scale, the structure shows a velocity gradient 
of $\sim$~2.5--3.0~$\km\ps$degree$^{-1}$ and has relatively enhanced \twCO\ emission, 
which is likely from MCs within the Outer Arm (or the Norma--Cygnus Arm) 
in the first quadrant.
Moreover, the large-scale structure is comprised of several 
interesting substructures in the $l$--$v$ space, 
which themselves are worthy of further investigation.
 
In addition to the enhanced \twCO\ emission, considerable molecular gas with
weaker CO emission at more negative LSR velocities is also discerned 
from Figure~\ref{f9} (see the regions between the red lines, i.e.,
$V_{\rm LSR} =-1.57 \times l+4.34 \pm12$~km~s$^{-1}$).
The most negative velocity is at $\sim -75$~km~s$^{-1}$ with
low surface brightness.
These MCs are probably in a distant section of the Scutum--Centaurus Arm 
\citep[designated as the OSC by][]{2011ApJ...734L..24D},
which is referred to as the EOG region by \cite{2017ApJS..230...17S}.
The trend of the molecular gas in the $l$--$v$ space
is approximately described by a linear fit of
$V_{\rm LSR} =-1.57 \times l+4.34$~km~s$^{-1}$, which is in good agreement with
previous studies \citep[see][]{2011ApJ...734L..24D,2017ApJS..230...17S}.
The velocity gradient of $\sim1.6 \km\ps$~degree$^{-1}$ for the EOG gas 
is smaller than that of the molecular gas of the Outer Arm in the $l$--$v$ space
\citep[e.g., $\sim2.8 \km\ps$~degree$^{-1}$ in][]{2016ApJ...828...59S}.
The uncertainty of the tentative fit is largely due to the limited longitude 
coverage. On the other hand, the OSC Arm may consist of 
several broken lines with somewhat different slopes in the $l$--$v$ space. 
Further MWISP data from $l=0^{\circ}$ to $26^{\circ}$ 
and $l=50^{\circ}$ to $100^{\circ}$ can give us a firmer conclusion
\citep[e.g., see Figure~3 in][]{2015ApJ...798L..27S}.

Figure~\ref{f10} displays the distribution of the EOG \twCO\ emission.
Samples in the figure are selected from the MCs
with $V_{\rm LSR} \lsim-1.57 \times l+4.34$~km~s$^{-1}$, 
which reduces possible contamination from the Outer Arm gas.
We find that these most distant MCs are quite sparse in the $l-b$ space and indeed
located above the $b=0^{\circ}$ plane. 
For regions of $l \sim 40^{\circ}-50^{\circ}$, the lack of the EOG CO emission 
in the map indicates that the slope of $-1.57$ in our fitting is probably 
too steep for the distant MCs at larger longitudes (e.g., a slightly flat slope of 
$\gsim -1.4$ for $l \gsim40^{\circ}$ CO gas in the $l$--$v$ space). 
Two intensity peaks are located at
$b\sim$1\fdg2 and $b\sim$2\fdg8,
or $\sim$370~pc and $\sim$870~pc above the $b=0^{\circ}$ plane, assuming
a median heliocentric distance of 17.8~kpc \citep{2017ApJS..230...17S}.
These values are several times larger than that of the Outer Arm gas \citep[e.g., 0\fdg42 
or $\sim$110 pc at a heliocentric distance of 15 kpc;][]{2016ApJ...828...59S},
indicating the different distributions 
of the two MC groups for the distant $V_{\rm LSR} < 0$~km~s$^{-1}$ gas.
Indeed, the more negative the value of the LSR velocity is,
the higher the Galactic latitude of the distant molecular gas.

\subsubsection{Bubble-like $C^{18}O$ Structure near the \HII\ region W40}
At similar distances,
there may be multiple MCs with different velocity fields in the ISM,
leading to complicated molecular gas structures.
Simultaneous \twCO, \thCO, and C$^{18}$O~($J$=1--0) line observations
can provide a more complete picture of the molecular gas properties, such as 
structures, kinematics, and dynamics.
Star formation is associated with molecular gas. 
The MWISP survey is helpful for studying the relationship between 
MCs and star-forming activity in the Milky Way.

An intriguing case is the W40 region, which is part of the 
Serpens$/$Aquila Rift MC complex. The region
is located at a distance of 436.0$\pm9.2$~pc according to the measurements 
of proper motions of seven stars across the MC complex \citep{2017ApJ...834..143O}.
\cite{2017ApJ...837..154N} covered the MC complex using
\twCO, \thCO, and C$^{18}$O~($J$=2--1) lines with the 1.85~m telescope.
The gas toward W40 within an area of $\sim$~1 square degree 
was further investigated by the 
Nobeyama 45~m telescope \citep{2018PASJ..tmp..131S}.
Based on the MWISP survey on the 13.7~m telescope, 
a prominent bubble-like structure is unveiled in the C$^{18}$O intensity map
(e.g., see the upper right corner of Figure~\ref{f7}).

Figure~\ref{f11} displays the close-up view toward the W40 region, where
the blue, green, and red colors represent the \twCO, \thCO, and C$^{18}$O 
emission, respectively. The circular morphology of the dense gas in the 
three-color image is centered at ($l=$29\fdg07, $b=$3\fdg82), with
a radius of 0\fdg45. Obviously, the emission of the dense gas
traced by \thCO\ and C$^{18}$O lines is mainly concentrated in the western half
of the bubble-like structure, while the C$^{18}$O structure in the eastern
half is broken. The LSR velocity of C$^{18}$O emission peaks at 
$V_{\rm LSR} \sim$7.5~km~s$^{-1}$ on the bubble-like structure, 
where both the \twCO\ and \thCO\ lines show the striking self-absorption features
(see the typical spectra to the right of the figure).

The strongest \twCO\ emission of $T_{\rm MB}\sim$37~K at 
$V_{\rm LSR} \sim$4.5~km~s$^{-1}$ is located at ($l=$28\fdg75, $b=$3\fdg51), 
which is close to the center of the \HII\ region W40 
(see the purple circle in Figure~\ref{f11}). 
The typical peak temperature of the optically thick \twCO\ line is nearly 10~K 
in some bright regions, while considerable \twCO\ emission has a lower
peak temperature of $\sim$4--6~K in other faint regions.
Huge regions of molecular gas with relatively weak CO emission extend 
several tens of square degrees over the FOV (e.g., see 0--10~km~s$^{-1}$ and 
10--20~km~s$^{-1}$ maps in Figure~\ref{f3a}).

Figure~\ref{f12} displays the gas distribution in velocity intervals of 
3--5, 5--7, 7--9, and 9--11~km~s$^{-1}$ toward the complex.
For the 3--5~km~s$^{-1}$ gas, the very strong CO emission is found to 
surround the W40 \HII\ region (the 9$'$ purple circle in the map), 
while CO emission seems to be lacking toward the center of the \HII\ region.
The enhanced dense gas, which is represented by the bright CO emission,
exhibits a bubble-like structure in the 5--7 and 7--9~km~s$^{-1}$ maps.
We use a large circle with a diameter of 0\fdg9 to show the 
bubble-like structure.

In the northeastern region of the broken C$^{18}$O bubble-like
structure (i.e., near
$l \sim$29\fdg5 and $b \sim$4\fdg0), several finger-like protrusions traced by 
\twCO\ and \thCO\ emission are found to point to the W40 \HII\ region
(see the $V_{\rm LSR} \sim$9--11~km~s$^{-1}$ map in Figure~\ref{f12}).
Moreover, the CO data reveal many interesting features, including
rim-bright MCs, molecular gas bubbles$/$cavities, and shells$/$arcs 
in the complicated region.
These features may be related to the strong feedback from the \HII\ region W40.

Figure~\ref{f13} exhibits the velocity distribution of the C$^{18}$O gas
toward the W40 region. Three main velocity components are recognized, such as 
the partial shell structure surrounding W40 at $V_{\rm LSR} \sim$~5--6~km~s$^{-1}$
(thick blue), the western half bubble-like structure at 
$V_{\rm LSR} \sim$~7--8~km~s$^{-1}$ (thin blue to green),
and the eastern broken bubble-like structure at $V_{\rm LSR} \sim$~9~km~s$^{-1}$ (red).
Some line-broadening features of \twCO\ gas (e.g., $V_{\rm LSR} \lsim$2~km~s$^{-1}$ and
$V_{\rm LSR} \sim$12~km~s$^{-1}$) are also revealed in the whole region,
indicating the potential outflow candidates.
Detailed analysis of the nearby region will be helpful for
understanding the relationship between the molecular gas and the ongoing star formation.


\subsubsection{CO gas distribution perpendicular to the Galactic plane}
\cite{1994ApJ...436L.173D} suggested that a thick molecular 
disk traced by faint CO emission is well above the central thin CO disk in
the inner Galaxy. Their results are based on CO observations of the 1.2~m telescope
for narrow regions of Galactic longitudes $l= 30^{\circ}, 40^{\circ}$, and $50^{\circ}$ 
and Galactic latitudes $|b| < 4^{\circ}$.
Benefiting from the unbiased, high-sensitivity, and large-area MWISP CO survey,
we have a pretty good chance of studying
the second, faint, and thick disk component of the molecular gas 
toward the inner Galaxy.

Channel maps of \twCO\ and \thCO\ (Figures~\ref{f3a} and \ref{f4a}) show that 
there are many MCs near the terminal velocity extending up to $\sim 1^{\circ}$--3$^{\circ}$
from the plane. Some MCs have extremely $b$ values of $>4^{\circ}$. 
These MCs at $|b| \gsim 1^{\circ}$ and $V_{\rm LSR} \geq 60$~$\km\ps$ are 
relatively isolated with respect to the molecular gas near the Galactic plane.
The MCs with weak CO emission, which are often associated with the enhanced 
\mbox{H\,\textsc{i}} emission 
\citep[e.g., to compare with the GALFA data,][]{2011ApJS..194...20P}, have 
little velocity crowding at regions of relatively high $b$.
 
Peak \twCO\ velocities of $V_{\rm LSR} \geq 60$~$\km\ps$ molecular gas
are extracted pixel by pixel for the whole region of $+$25\fdg8$\lsim l \lsim +$49\fdg7 
and $|b|\lsim$5\fdg2.
The pixels with three consecutive channels $\geq 3\times$~rms 
(or $\sim$0.8~K for the smoothed
velocity resolution of 0.5~km~s$^{-1}$) would be considered to be the valid
signal. The near kinematic distances of the CO signal are calculated assuming
the flat Galactic rotation curve model \citep[e.g.,][]{2014ApJ...783..130R}. 
Then, the distances from the $b= 0^{\circ}$
plane (i.e., $z$) are estimated for each pixel.
To reduce sample fluctuations, $z$ values are combined 
every 10~pc per bin from $-500$~pc to $+$400~pc.
CO intensities are averaged for all valid CO signal per bin.
The error bars are estimated from the express of 
$I_{\rm mean}$(CO)/(number of pixels per bin)$^{0.5}$.
As a consequence, the error bars of the high-$z$ samples are obviously larger than those of
points near the Galactic plane. 

Figure~\ref{f14} displays the vertical distribution of the CO gas.
We find that one Gaussian model produces a poor fit to the histogram of
the CO distribution from the Galactic plane (e.g., $\chi^{2}$/dof=94.8/86).
The full width at half maximum (FWHM) of the one Gaussian component is 162.7~pc, 
which is larger than results from other studies \citep[e.g., FWHM$\sim$~90--110~pc for
the inner Galaxy, e.g.,][]{1994ApJ...433..687M,2006PASJ...58..847N,
2016PASJ...68....5N,2015ARA&A..53..583H,2016ApJ...818..144R}. 
Moreover, the mean emission peak at $z \sim -11$~pc and the prominent broad wing at 
$|z| \gsim 100$~pc cannot be fitted by the one Gaussian model. 

On the contrary, the model of two Gaussian components works better
($\chi^{2}$/dof=60.5/84, also see the red-thick line in Figure~\ref{f14}). 
Table~2 shows the parameters of the best fit.
The best fit of the zero-point is at $\sim$~3.8~K~$\km\ps$, which is roughly 
5$\sigma$ of the CO integrated intensity. The value of the zero-point depends
on the current rms level of the MWISP survey, in which
samples with three consecutive channels greater than $3\times$~rms 
($\sim 0.8$~K) are chosen as the valid \twCO\ signal of the MCs.
A few points near the zero-point
are from the noise (e.g., bad channels and baseline fluctuations in
the velocity axis of the 3D data cubes).
These bad points, which are randomly distributed in the area, 
are readily discerned by checking their spectra. 

Samples away from the Galactic plane with values above 3.8~K~$\km\ps$
are actually from the high-$z$ CO emission. An example is the unusual peak
at $z \sim -435$~pc (Figure~\ref{f14}). 
CO emission of this feature is mostly from MC G40.331$-$4.302
at $V_{\rm LSR} \sim$77--84~km~s$^{-1}$. 
The high $|b|$ MC is associated with the W50 nebula, 
which has a near kinematic distance of 4.9~kpc, 
as estimated from the large-scale gas at $V_{\rm LSR} =$77~km~s$^{-1}$
\citep[see the detailed \mbox{H\,\textsc{i}} and CO analysis in][]{2018ApJ...863..103S}.
At the distance of 4.9~kpc, the true displacement from the $b=0^{\circ}$ plane is 
$z \approx -$370~pc for the $b=-$4\fdg302 molecular gas,
which is likely related to the energetic jets of the microquasar SS~433 in W50 
\citep[i.e., jet--ISM interactions discussed in][]{2018ApJ...863..103S}.

The two Gaussian components correspond to the central thin molecular gas disk 
(FWHM=88.5~pc) and the extended thick CO disk (FWHM=276.8~pc), respectively.
The thickness ratio of the two Gaussian components is 
FWHM(Thick plane)/FWHM(Thin plane)$\sim$3.1
for the region of Galactocentric distances of $R_{\rm GC} \sim$~3.9--6.4~kpc 
(mean at $\sim$~5.4~kpc for all samples) 
toward the inner Galaxy. 
These results are consistent with 
previous studies \citep{1994ApJ...436L.173D,1994ApJ...437..194M}.
The total intensity ratio between $|z| \gsim 100$~pc and
$|z| \lsim 100$~pc is about 0.03, indicating very
faint CO emission for the second (or thick) molecular gas component.

In the above calculation, we used the A5 rotation curve model of \cite{2014ApJ...783..130R},
in which the values of $R_0$=8.34~kpc and $V_0$=240~km~s$^{-1}$ are adopted.
The thickness of the molecular gas disk will increase by a factor of $\sim$1.1 for
the parameters of $R_0$=8.5~kpc and $V_0$=220~km~s$^{-1}$.

We have shown that CO emission in the first quadrant of the inner Galaxy 
is somewhat below the Galactic plane based on channel maps in Figure~\ref{f3a}.
Further quantitative analysis indicates that the distribution of molecular gas is 
actually below the $b =0^{\circ}$ plane for the $V_{\rm LSR} \gsim$~60~km~s$^{-1}$ 
CO emission (Figure~\ref{f14} and Table~2).
Assuming that the inner disk of our Galaxy is flat and tilted, we derive
the Sun's offset from the Galactic physical midplane based on
the molecular gas distribution along the Galactic latitude. Considering 
$\frac{z_{\rm Sun}}{R_0} \approx \frac{z_{\rm peak}}{R_{\rm GC}(\rm mean)}$,
the Sun is about $8.34 \times \frac{11.3}{5.4}\approx$~17.5~pc 
above the physical midplane of the Milky Way, which agrees well with 
our previous study of $\sim$17.1~pc \citep{2016ApJ...828...59S}.

\section{Summary and Future Prospects}
The MWISP project is a large, systematic, and unbiased 
CO survey of the northern Galactic plane using the 13.7~m 
millimeter-wavelength telescope with a 3$\times$3 beam array.
As part of the legacy survey, the $\sim$250~deg$^2$ regions of the 
first quadrant of $l=+$25\fdg8 to $+$49\fdg7 and $|b|\lsim$5\fdg2
are completely mapped during the period of 2011--2018
in the \twCO, \thCO, and C$^{18}$O~($J$=1--0) lines.
Using the OTF mode, we achieved the high-quality CO data
with an angular resolution of $\sim50''$ and grid sampling of $30''$.
The rms noise level of the data is $\sim$0.5~K for \twCO\ 
at a channel width of 0.16~$\km\ps$ and
$\sim$0.3~K for \thCO\ and C$^{18}$O at 0.17~$\km\ps$.

Using the new CO data, we investigate the distant molecular gas 
in the first quadrant of the Milky Way. The \twCO\ distribution of the
most distant gas, which is weak and isolated in space and velocity,
is described by the equation of
$V_{\rm LSR} \lsim-1.57 \times l+4.34$~km~s$^{-1}$. We believe that
these MCs are from the Scutum--Centaurus Arm in the 
outer Galaxy, which are mainly concentrated in $b \gsim0^{\circ}$ regions 
of the first quadrant of the Milky Way due to
the warping of the Galactic outer disk.

A bubble-like structure with a diameter of 0\fdg9 in C$^{18}$O 
emission is revealed toward the Serpens$/$Aquila Rift MC complex. 
We suggest that the interesting structure, together with the surrounding
rim-bright MCs and molecular shells$/$arcs, may be related to the nearby 
\HII\ region W40, which is just located at the southwestern edge of the 
bubble-like dense gas structure. 

In addition to a thin CO disk with an FWHM of $\sim$88.5~pc, another faint and thick 
molecular gas disk with an FWHM of $\sim$276.8~pc is confirmed based on
the high-quality CO data. The thickness (or FWHM) of the faint CO component, 
which is about 3.1 times as wide as the thin CO layer, is roughly
comparable with the \mbox{H\,\textsc{i}} thickness for regions of $R_{\rm GC} \sim$~4--6~kpc
\citep[FWHM$\sim$~250--300~pc in][]{1990ARA&A..28..215D,1991ApJ...382..182L,
2016PASJ...68....5N}.
The second component of the faint CO gas is probably related to the activity 
of massive stars near the Galactic plane, which needs to be further explored.
The Sun's offset is found to be $\sim$17.5~pc above the Galactic disk,
according to the CO intensity peak at $z_{0}=-11.3$~pc.

\begin{figure}[ht]
\includegraphics[trim=50mm 0mm 0mm 0mm,scale=0.4,angle=0]{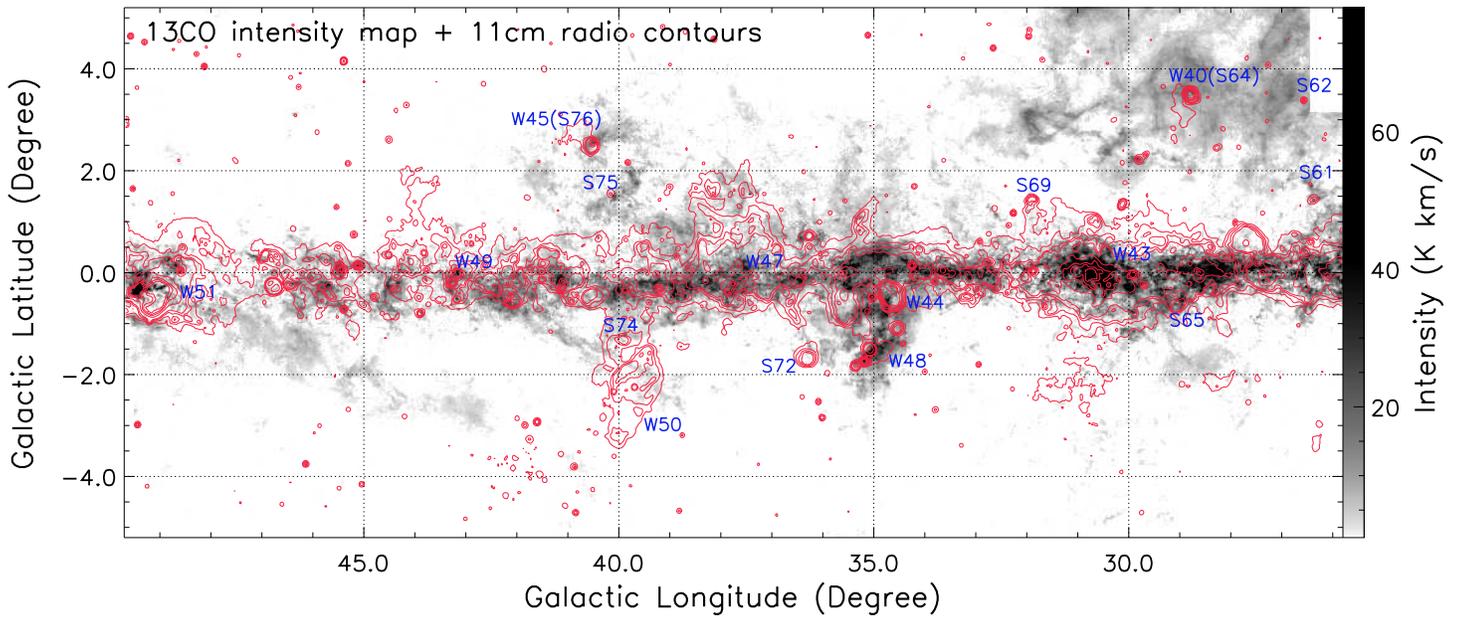}
\caption{
MWISP \thCO\ ($J$=1--0) intensity map in the 0--130~km~s$^{-1}$
interval, overlaid with the Effelsberg 11~cm radio contours
(250, 500, 750, 1$\times10^3$, 2.5$\times10^3$, 5$\times10^3$, 7.5$\times10^3$, and
1$\times10^4$~mK) from
\cite{1990A&AS...85..633R}. Some bright and/or extended radio sources are labeled
on the map.
\label{guidemap}}
\end{figure}

As shown by the examples in the paper, the unbiased MWISP CO survey is powerful for
studying the distribution and properties of the molecular gas,
the Galactic structures traced by MCs, and the relationship between
molecular gas and star formation.
According to the guide map of Figure~\ref{guidemap}, we exhibit
that the molecular gas is tightly correlated with the radio emission.
The bright radio emission is thought to be from the stellar feedback,
such as \HII\ regions and SNRs, which have profound effects on
their surrounding ISM. Therefore, the large-scale CO survey with 
high dynamic range is also an excellent dataset for studying
the effects of stellar feedback on MCs 
(i.e., outflows, \HII\ regions, stellar winds, and SNRs on the 
surrounding molecular gas environment).

Further MC identification is important for the extended CO emission
with velocity crowding. New methods, such as SCIMES
\citep[i.e., Spectral Clustering for Interstellar Molecular Emission Segmentation,][]
{2015MNRAS.454.2067C}, may be useful for defining the structures of the diffuse 
CO emission and assessing the global properties of the MCs.
\cite{2017ApJ...834...57M} recently investigated the properties of 8107 Galactic MCs
using a hierarchical cluster identification applied to the result of a Gaussian 
decomposition of the CfA 1.2~m CO data.
On the other hand, \cite{2016ApJ...818..144R} examined the spatial 
distribution of the diffuse gas traced by \twCO\ emission and the dense gas traced by 
\thCO\ and CS emission in the inner Galaxy.
Based on CO data of multiple surveys, they showed that CO emission is very important
for studying the dense star-forming MCs and the diffuse molecular ISM in the Galaxy.

Using the \twCO, \thCO, and C$^{18}$O~($J$=1--0) lines 
in combination with the wide velocity coverage,
the high-velocity resolution, and the large-scale mapping provided by the MWISP,
we can separate distinct clouds at similar velocities,
investigate detailed kinematic information of the molecular gas at complicated
regions, and provide the MC catalog traced by CO emission.
More sophisticated methods
will be explored to systematically analyze the spatial and 
kinematic features of MCs in the Milky Way based on the new CO survey.

\acknowledgments
We gratefully acknowledge the staff members of the Qinghai Radio
Observing Station at Delingha for their support of the observations.
We thank the anonymous referee for a careful reading of the manuscript
and several critical suggestions that improved the paper.
This work is funded by the National Key R\&D Program of China
through grants 2017YFA0402701 and 2017YFA0402702.
J.Y. acknowledges CAS support through grant QYZDJ-SSW-SLH047.
X.C. acknowledges support by the NSFC through grant 11473069.
Y.S. was supported by the NSFC through grant 11773077.

\facility{PMO 13.7m}
\software{GILDAS/CLASS \citep{2005sf2a.conf..721P}} 

\bibliographystyle{aasjournal}
\bibliography{references}

\begin{deluxetable}{ccc}
\tabletypesize{\scriptsize}
\tablecaption{Parameters of the MWISP Survey Data}
\tablehead{
\colhead{\begin{tabular}{c}
Molecular lines   	        \\  HPBW                     \\  Grid spacing                 \\  
Velocity separation   		\\  System temperature       \\  rms ($T_{\rm MB}$)    \\  
Mapping regions			\\
\end{tabular}} &
\colhead{\begin{tabular}{c}
\twCO~($J$=1--0)   		\\  $\sim49''$  	      \\   $30''$  		      \\  
0.16~km~s$^{-1}$  		\\  $\sim250$~K		      \\  $\sim0.5$~K		      \\
25\fdg8$\lsim l \lsim$49\fdg7, $\left | b \right | \lsim$5\fdg2      \\
\end{tabular}} &
\colhead{\begin{tabular}{c}
\thCO\ and C$^{18}$O~($J$=1--0)  \\ $\sim52''$   	      \\   $30''$                     \\
0.17~km~s$^{-1}$               \\  $\sim140$~K               \\  $\sim0.3$~K               \\
25\fdg8$\lsim l \lsim$49\fdg7, $\left | b \right | \lsim$5\fdg2      \\
\end{tabular}}
}
\startdata
\enddata
\end{deluxetable}

\begin{deluxetable}{ccccccc}
\tabletypesize{\scriptsize}
\tablecaption{Two Gaussian Components for the Inner Galactic CO Plane}
\tablehead{
\colhead{}  & \colhead{Narrow}  &  \colhead{}  &    \colhead{}  &   \colhead{Broad}  &   \colhead{}  & \colhead{}  \\
\cline{1-3}  \cline{5-6}
\colhead{$z_{0}$ (pc)}  & \colhead{FWHM (pc)}  &  \colhead{$I_{0}$ (K~$\km\ps$)}  &    
& \colhead{FWHM (pc)}  &  \colhead{$I_{0}$ (K~$\km\ps$)}  &   \colhead{$\chi^{2}$/dof}
} 
\startdata  
$-11.3$  &  88.5   &    40.9   &    &   276.8     &   20.8   &  60.5/84\tablenotemark{a}      \\
\enddata
\tablenotetext{a}{The residual is calculated with a weight of the mean CO intensity per bin.}
\tablecomments{$z_{0}$ is the displacement of the CO gas from the Galactic plane of $b=0^{\circ}$; FWHM is the thickness
of the CO layer; $I_0$ is the peak of the best fit. Note that the zero-point is at 3.8~K~$\km\ps$,
i.e., $I_{0}$ = $I_{\rm peak}$ = $I(z_0)$ + 3.8~K~$\km\ps$ (see Figure~\ref{f14} and
the text). Note that the $z_0$ value of the broad component is fixed to that of 
the narrow one. Samples are from the CO emission near the tangent point, 
i.e., $V_{\rm LSR} >$~60~km~s$^{-1}$.}
\end{deluxetable}

\begin{figure}[ht]
\includegraphics[trim=0mm 0mm 0mm 20mm,scale=0.31,angle=0]{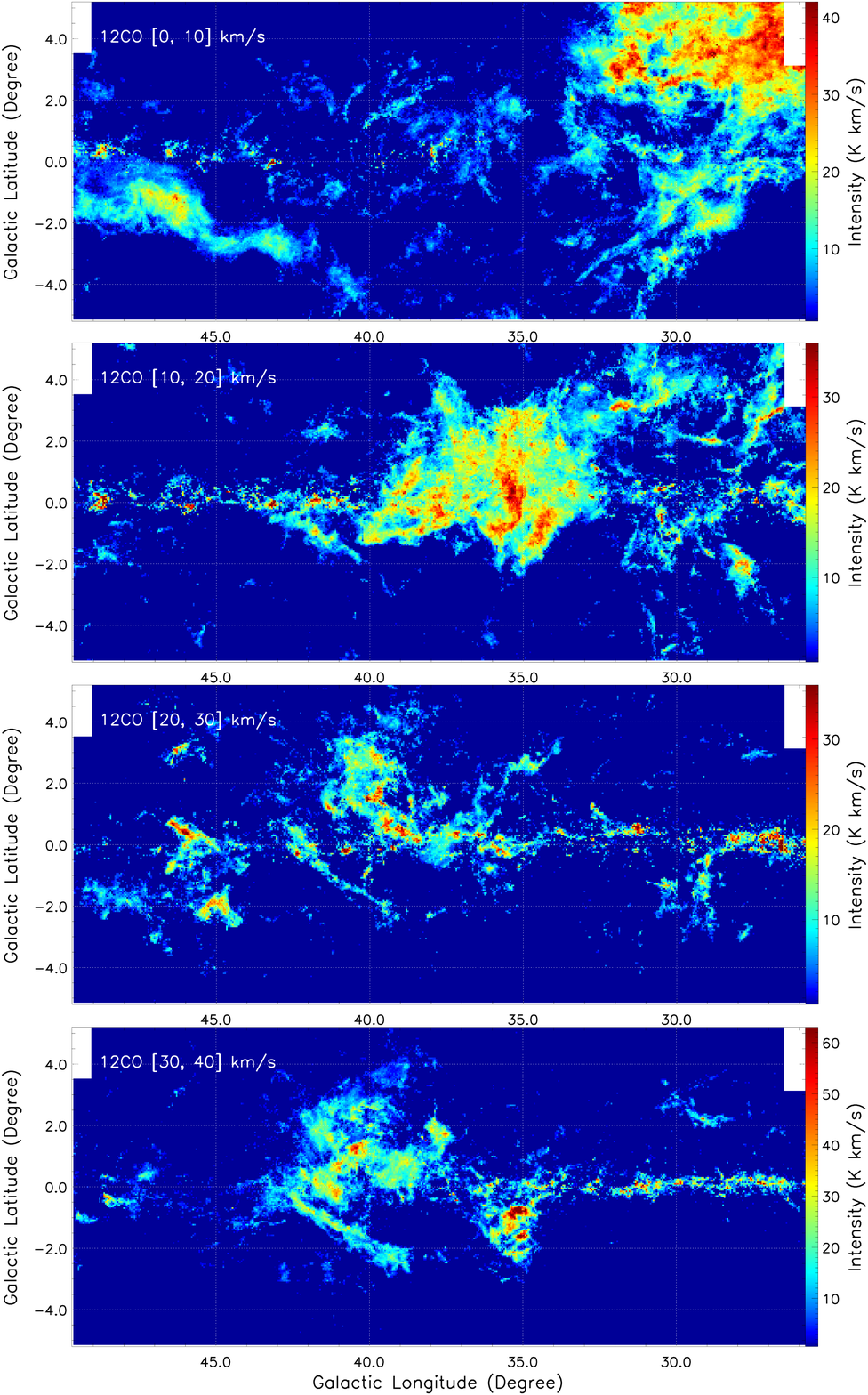}
\caption{
Channel maps of \twCO\ with velocity intervals of 10~km~s$^{-1}$
for $V_{\rm LSR}\geq$0~km~s$^{-1}$ gas.
\label{f3a}}
\end{figure}

\addtocounter{figure}{-1}
\begin{figure}[ht]
\includegraphics[trim=0mm 0mm 0mm 20mm,scale=0.31,angle=0]{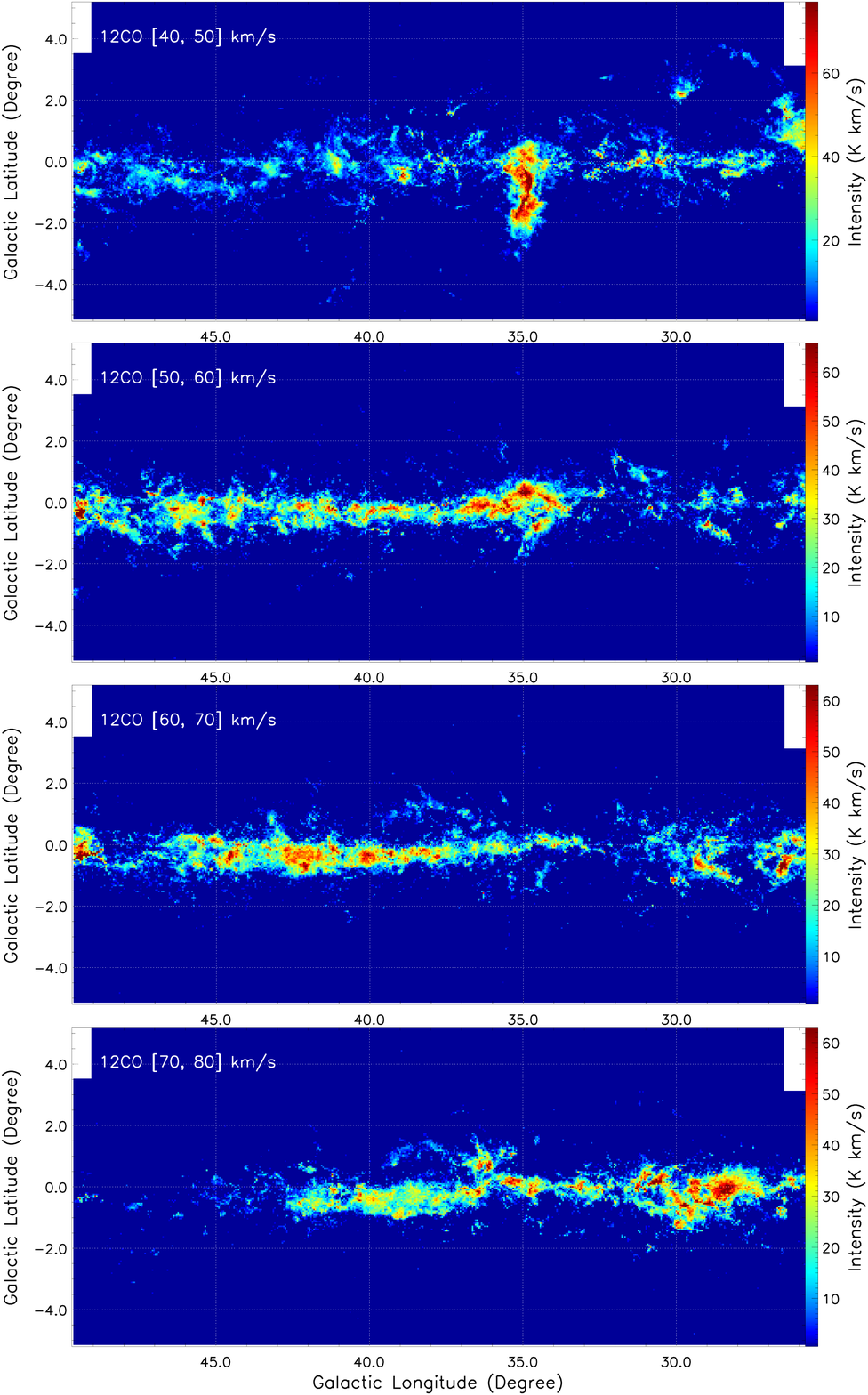}
\caption{
(Continued)
\label{f3b}}
\end{figure}

\addtocounter{figure}{-1}
\begin{figure}[ht]
\includegraphics[trim=0mm 0mm 0mm 20mm,scale=0.31,angle=0]{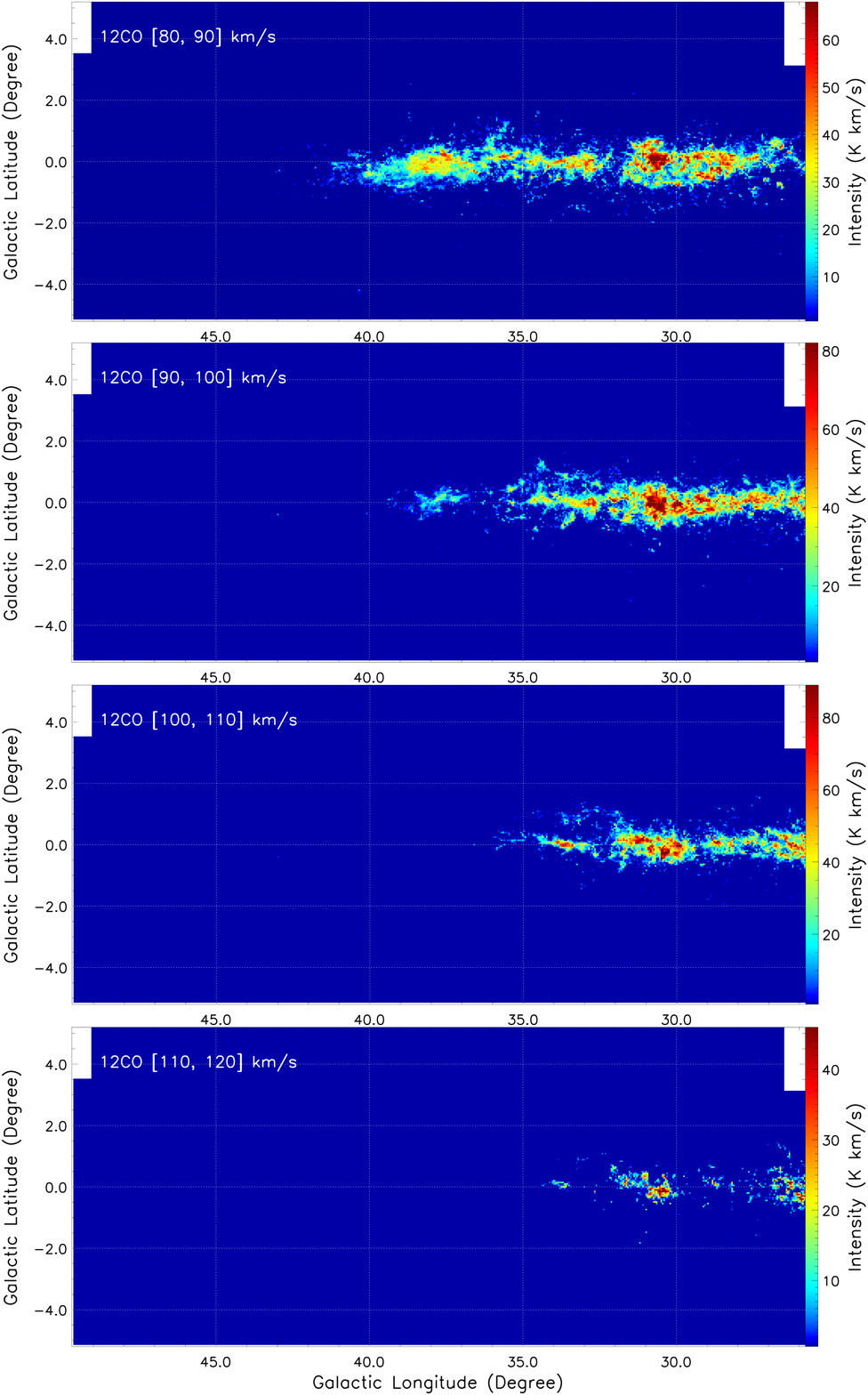}
\caption{
(Continued)
\label{f3c}}
\end{figure}

\begin{figure}[ht]
\includegraphics[trim=0mm 0mm 0mm 20mm,scale=0.31,angle=0]{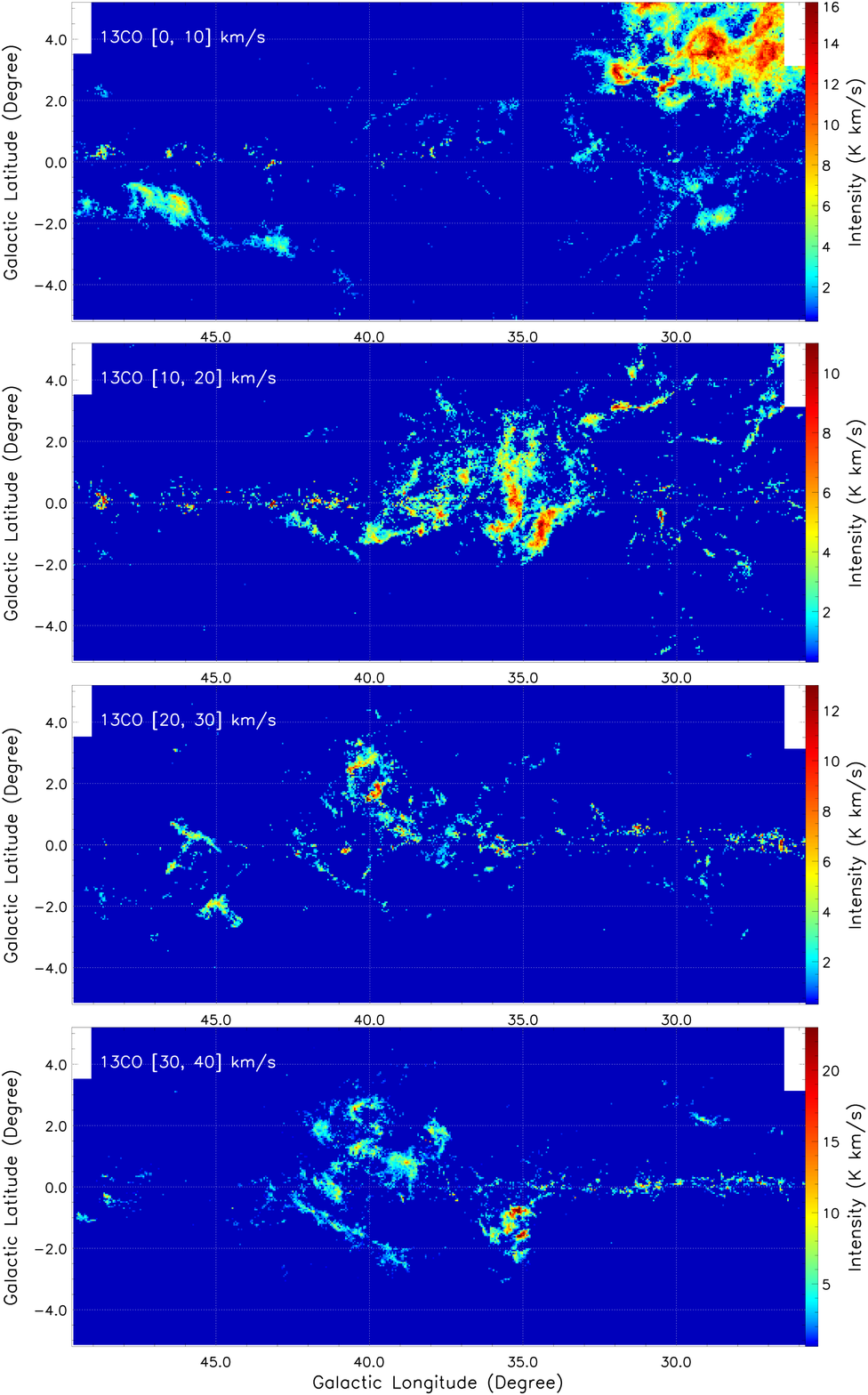}
\caption{
Channel maps of \thCO\ with velocity intervals of 10~km~s$^{-1}$
for $V_{\rm LSR}\geq$0~km~s$^{-1}$ gas.
\label{f4a}}
\end{figure}

\addtocounter{figure}{-1}
\begin{figure}[ht]
\includegraphics[trim=0mm 0mm 0mm 20mm,scale=0.31,angle=0]{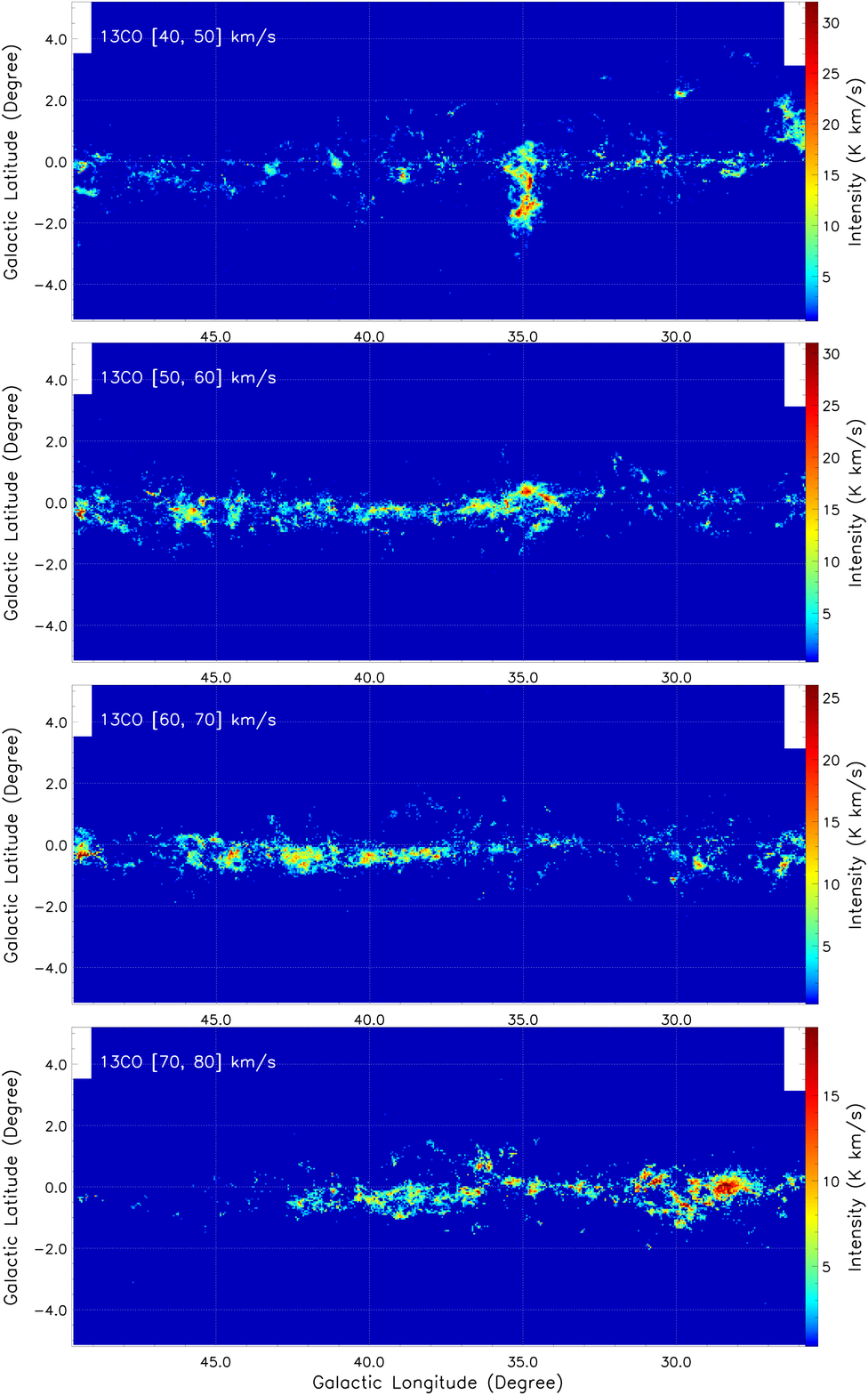}
\caption{
(Continued)
\label{f4b}}
\end{figure}

\addtocounter{figure}{-1}
\begin{figure}[ht]
\includegraphics[trim=0mm 0mm 0mm 20mm,scale=0.31,angle=0]{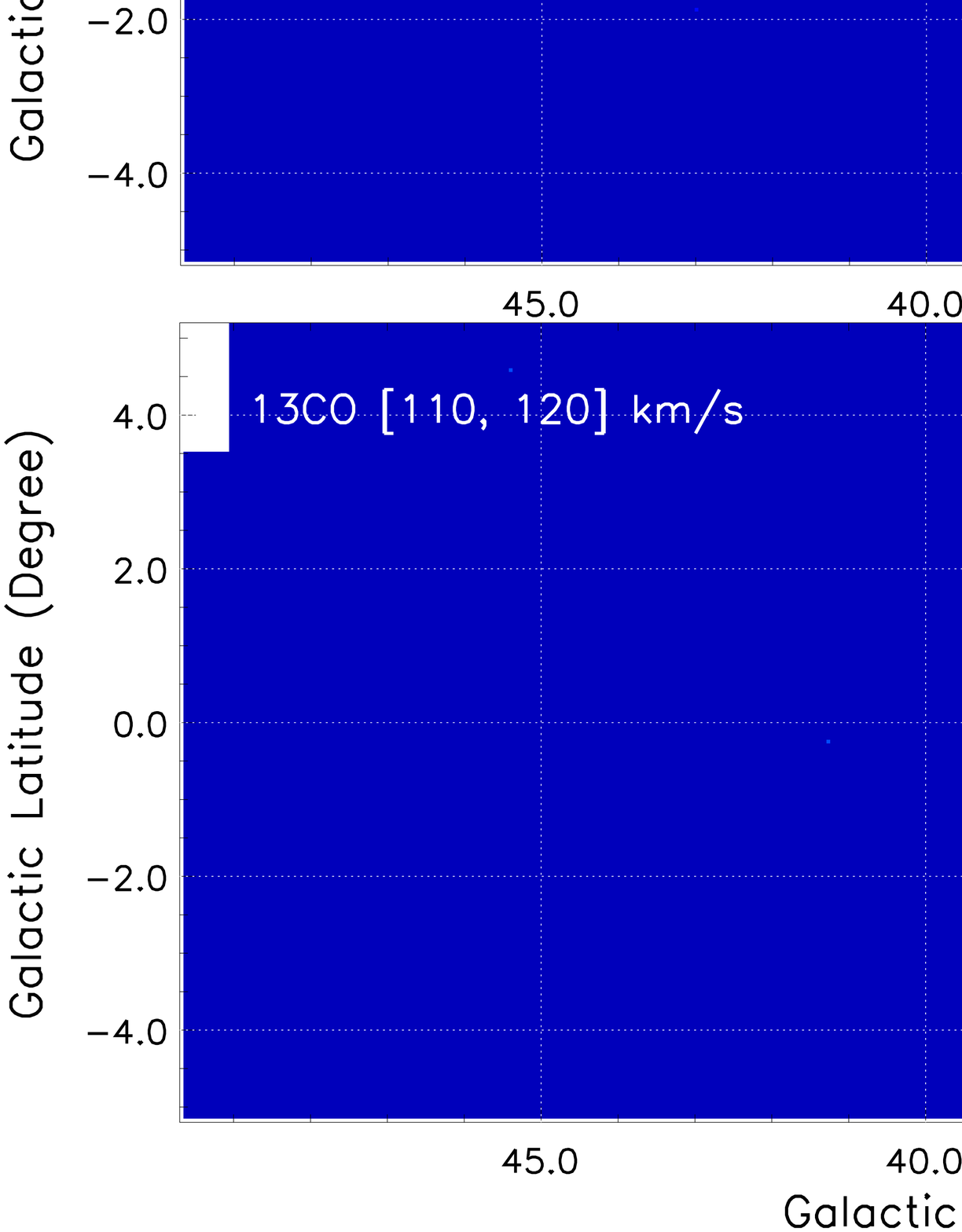}
\caption{
(Continued)
\label{f4c}}
\end{figure}

\begin{figure}
\gridline{
          \hspace{-8ex}
          \fig{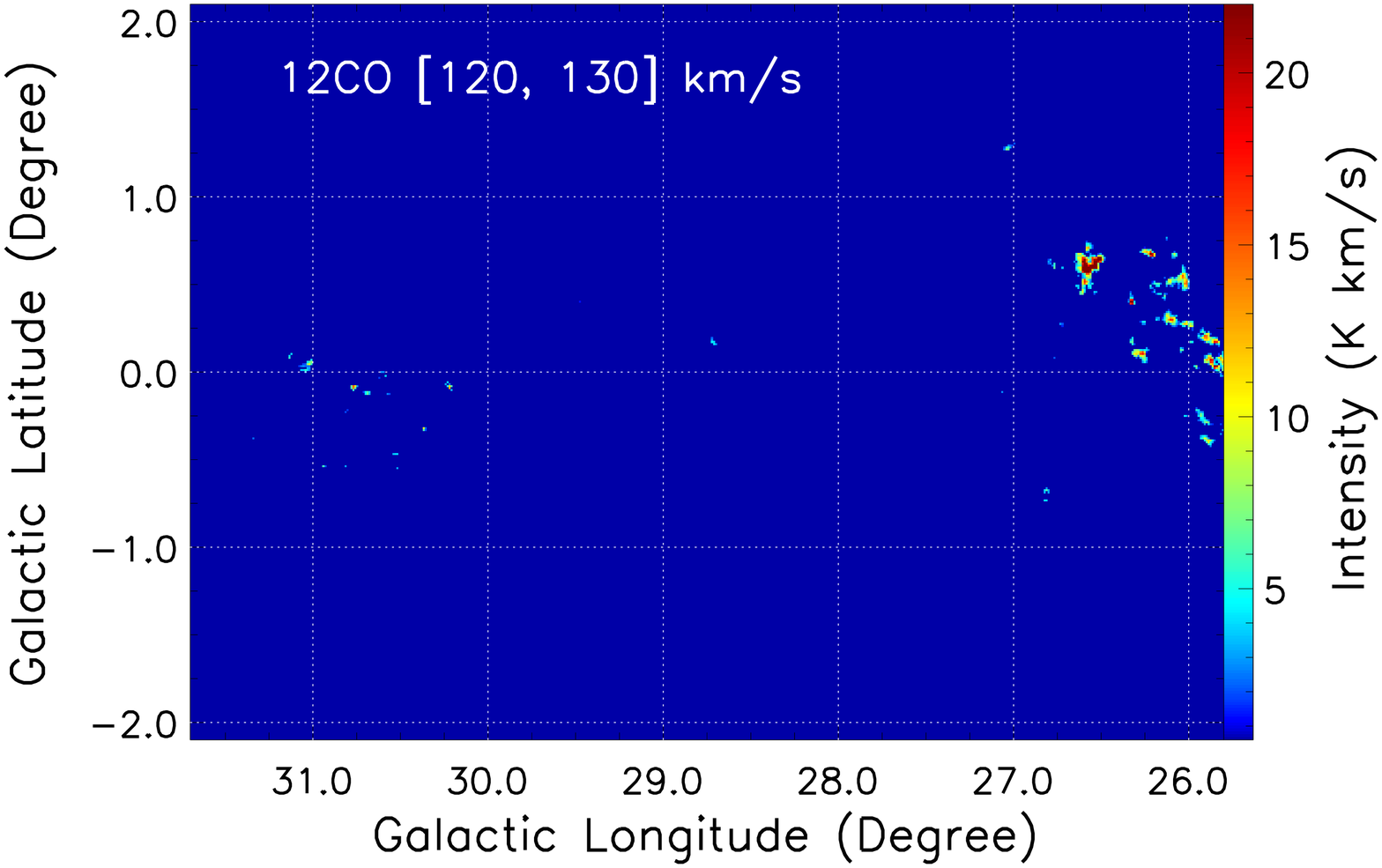}{0.6\textwidth}{}
          \hspace{-2ex}
          \fig{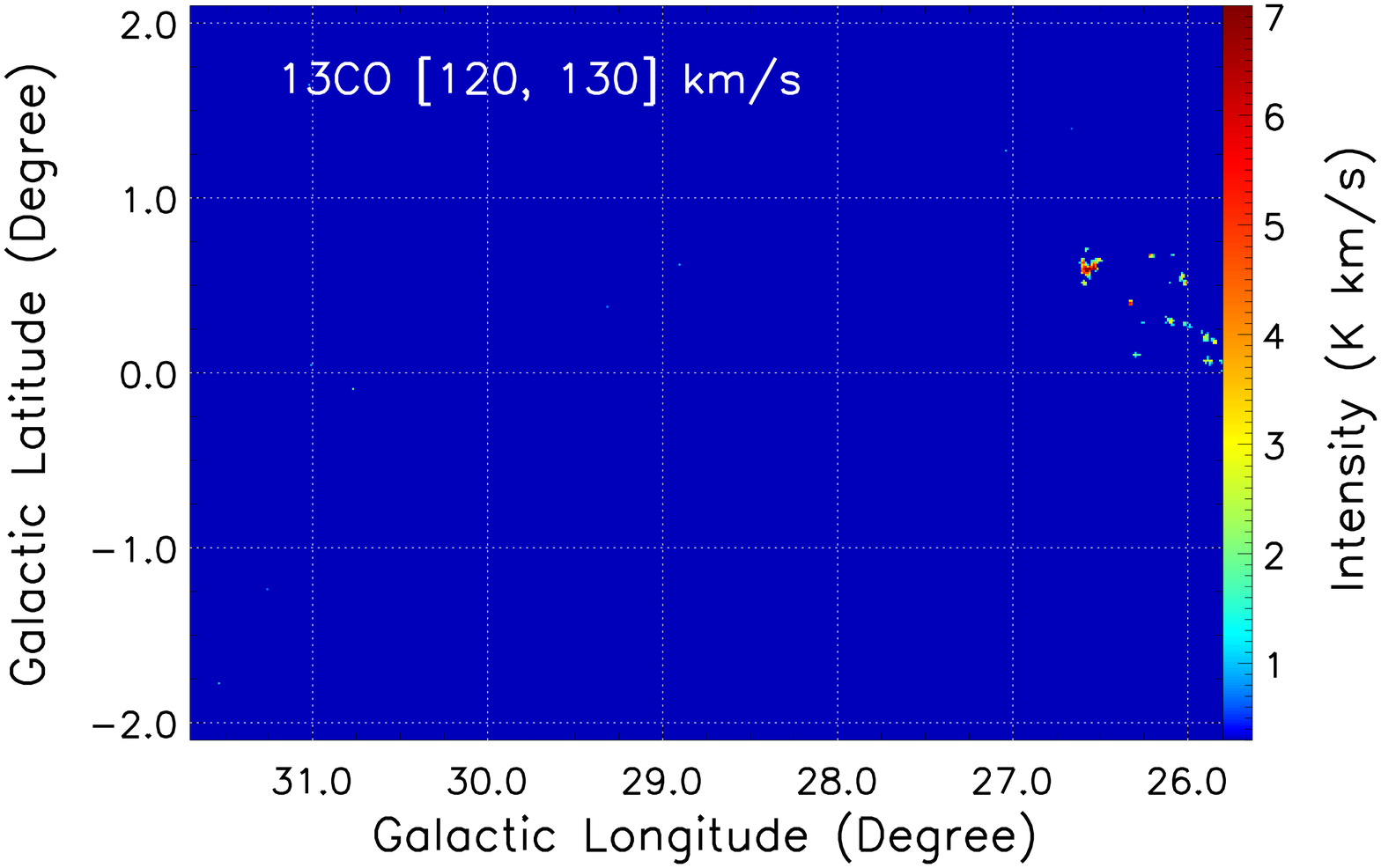}{0.6\textwidth}{}
          }
\caption{
Left panel: 
The integrated emission of the \twCO\ emission
in the interval of $120$ to 130~km~s$^{-1}$.
Right panel: 
The integrated emission of the \thCO\ emission
in the same interval. 
\label{f5}}
\end{figure}

\begin{figure}
\gridline{
          \hspace{-10ex}
	  \fig{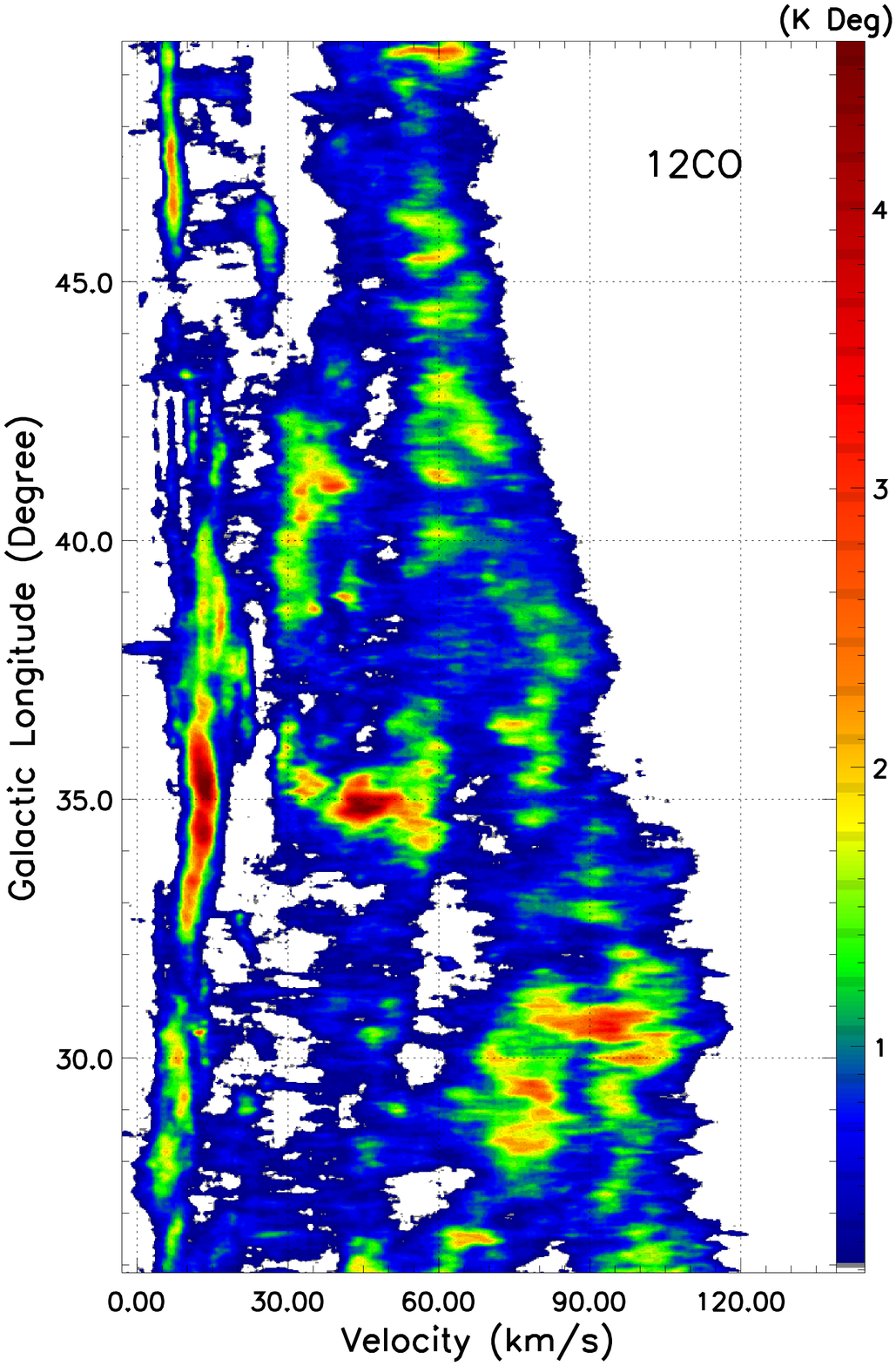}{0.65\textwidth}{}
	  \hspace{-5ex}
          \fig{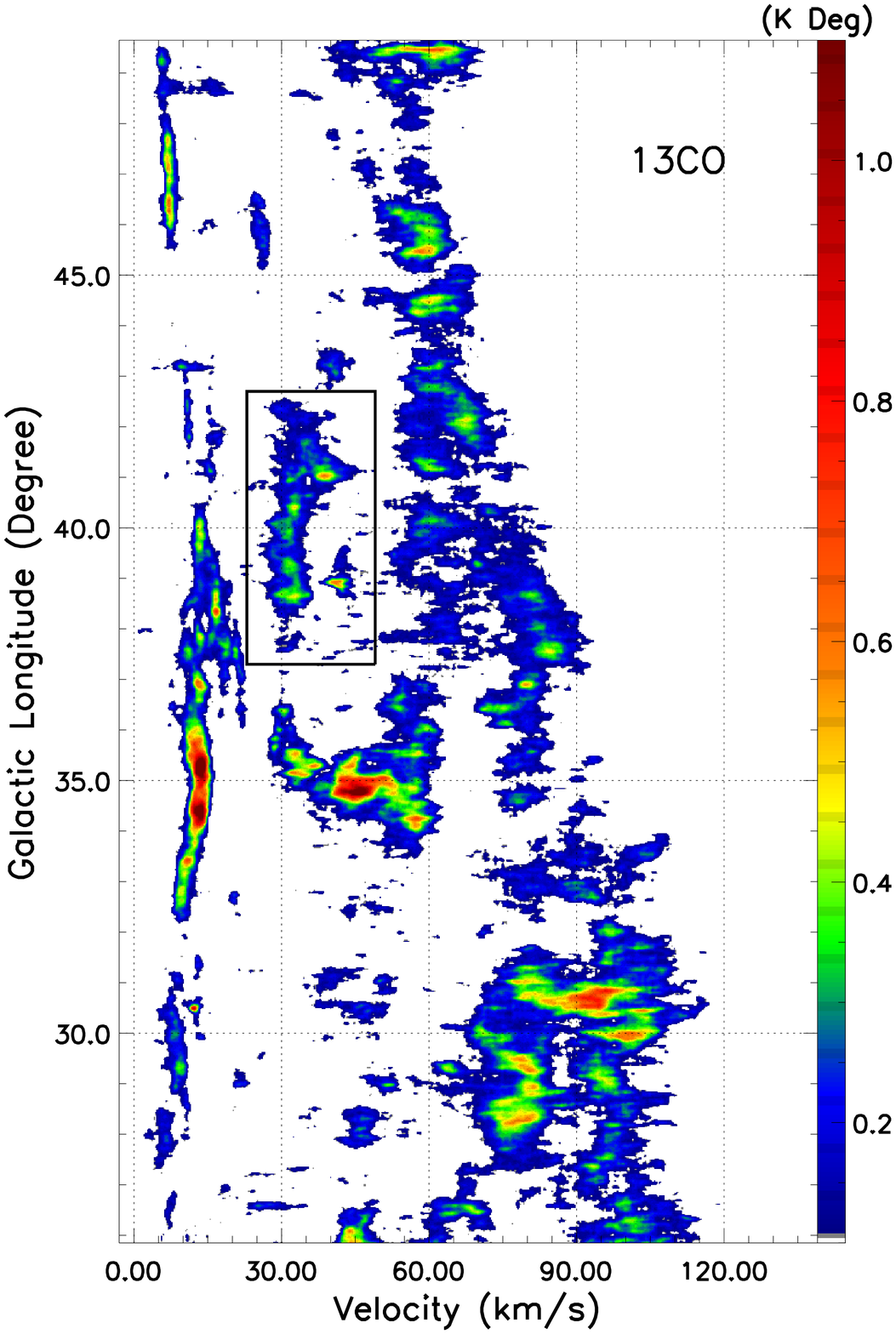}{0.65\textwidth}{}
          }
\caption{
Longitude--velocity diagrams of the \twCO\ and \thCO\ emission
for $V_{\rm LSR}\geq$0~km~s$^{-1}$ gas.
The rectangle in the right panel shows the region of
the dense molecular gas associated with
\mbox{H\,\textsc{ii}} regions Sh~2-75 and Sh~2-76,
which was not fully covered in latitude by the GRS survey \citep{2006ApJS..163..145J}.
\label{f6}}
\end{figure}

\begin{figure}[ht]
\includegraphics[trim=25mm 0mm 0mm 0mm,scale=0.7,angle=0]{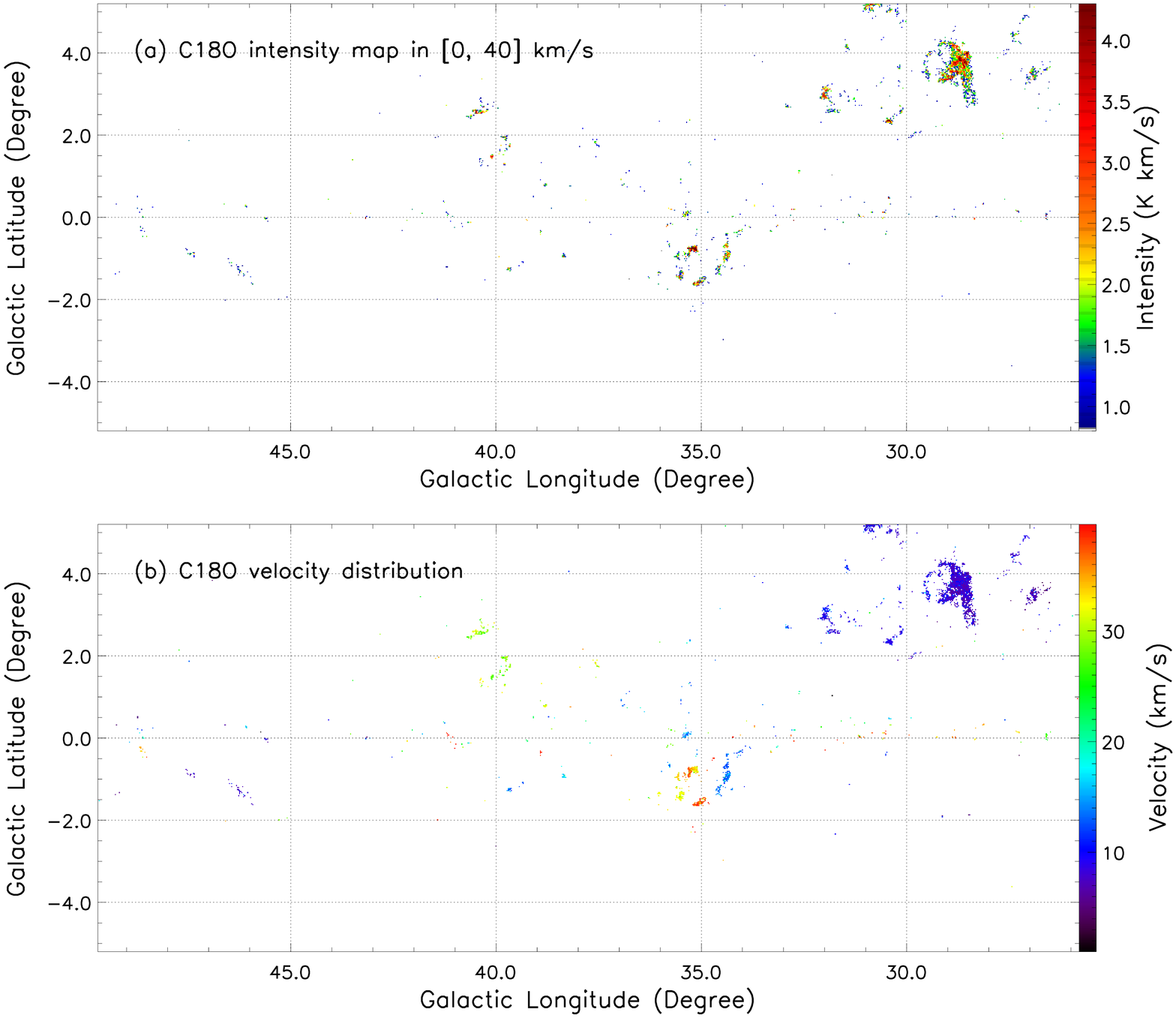}
\caption{
Intensity map and velocity distribution of C$^{18}$O emission for
$V_{\rm LSR}$=0--40~km~s$^{-1}$ gas.
\label{f7}}
\end{figure}

\begin{figure}[ht]
\includegraphics[trim=20mm 0mm 0mm 0mm,scale=0.9,angle=0]{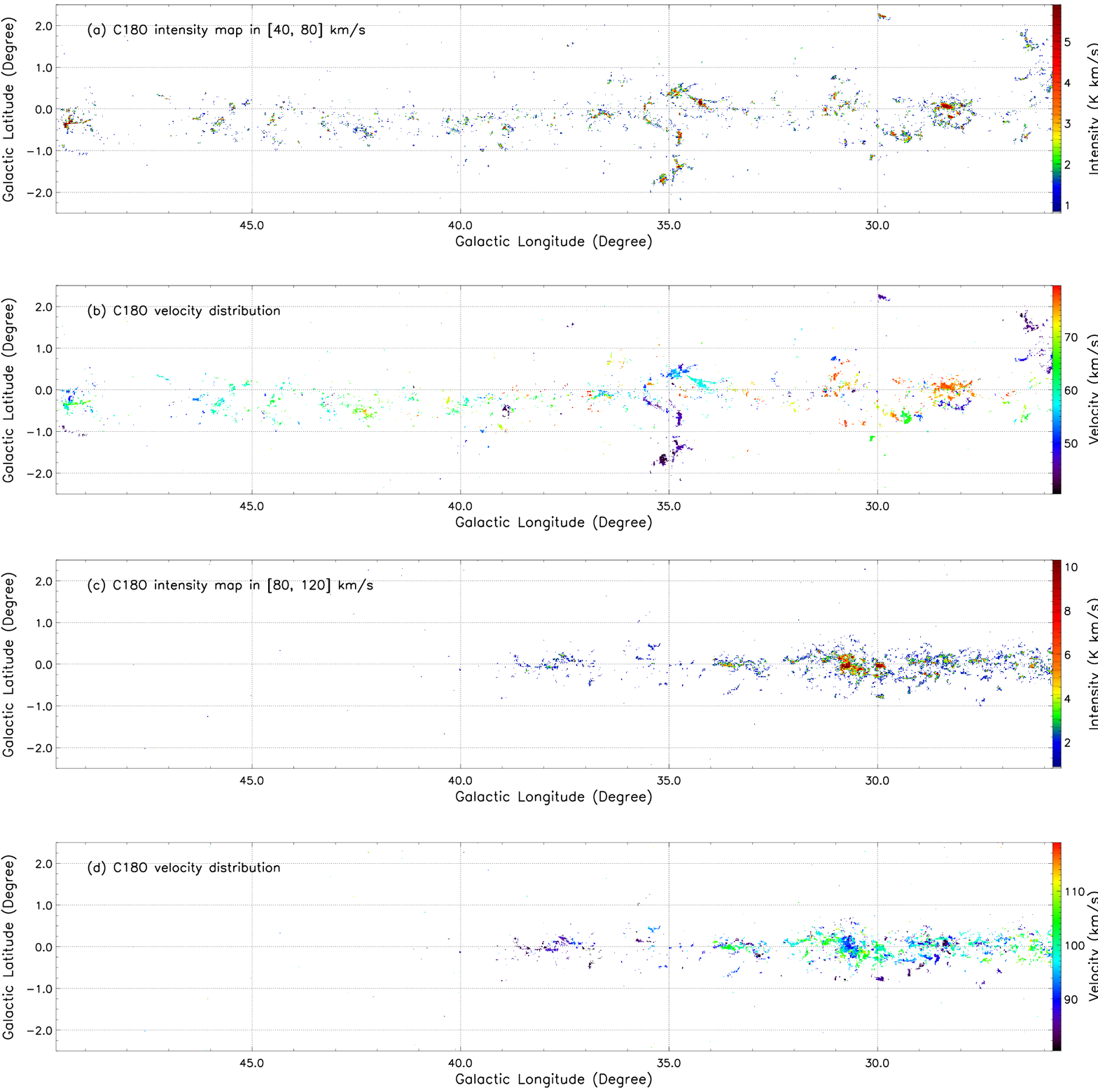}
\caption{
Intensity map and velocity distribution of C$^{18}$O emission for
$V_{\rm LSR}$=40--80~km~s$^{-1}$ and 80--120~km~s$^{-1}$ gas.
\label{f8}}
\end{figure}

\begin{figure}[ht]
\includegraphics[trim=15mm 0mm 0mm 5mm,scale=0.8,angle=0]{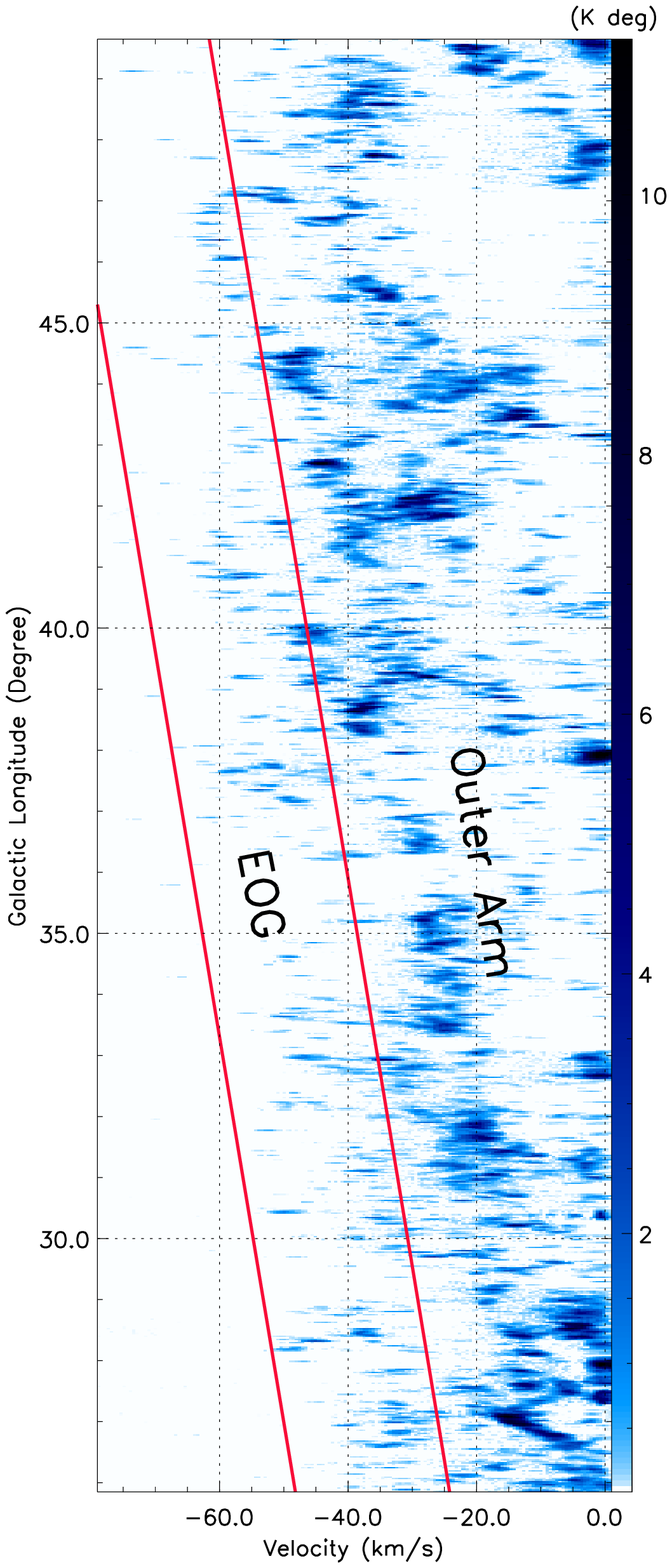}
\caption{
Longitude--velocity diagram of the \twCO\ emission
for $V_{\rm LSR}\lsim$0~km~s$^{-1}$ gas. The region between the red lines
($V_{\rm LSR} =-1.57 \times l+4.34 \pm12$~km~s$^{-1}$) 
contains most of the detected CO gas in the EOG. 
The Outer Arm and the EOG regions are labeled on the map.
Note that the \twCO\ signal is multiplied by a factor of 100 
for the corresponding velocity range due to the 
weak emission of the distant MCs (see the text).
\label{f9}}
\end{figure}

\begin{figure}[ht]
\includegraphics[trim=25mm 0mm 0mm 0mm,scale=0.36,angle=0]{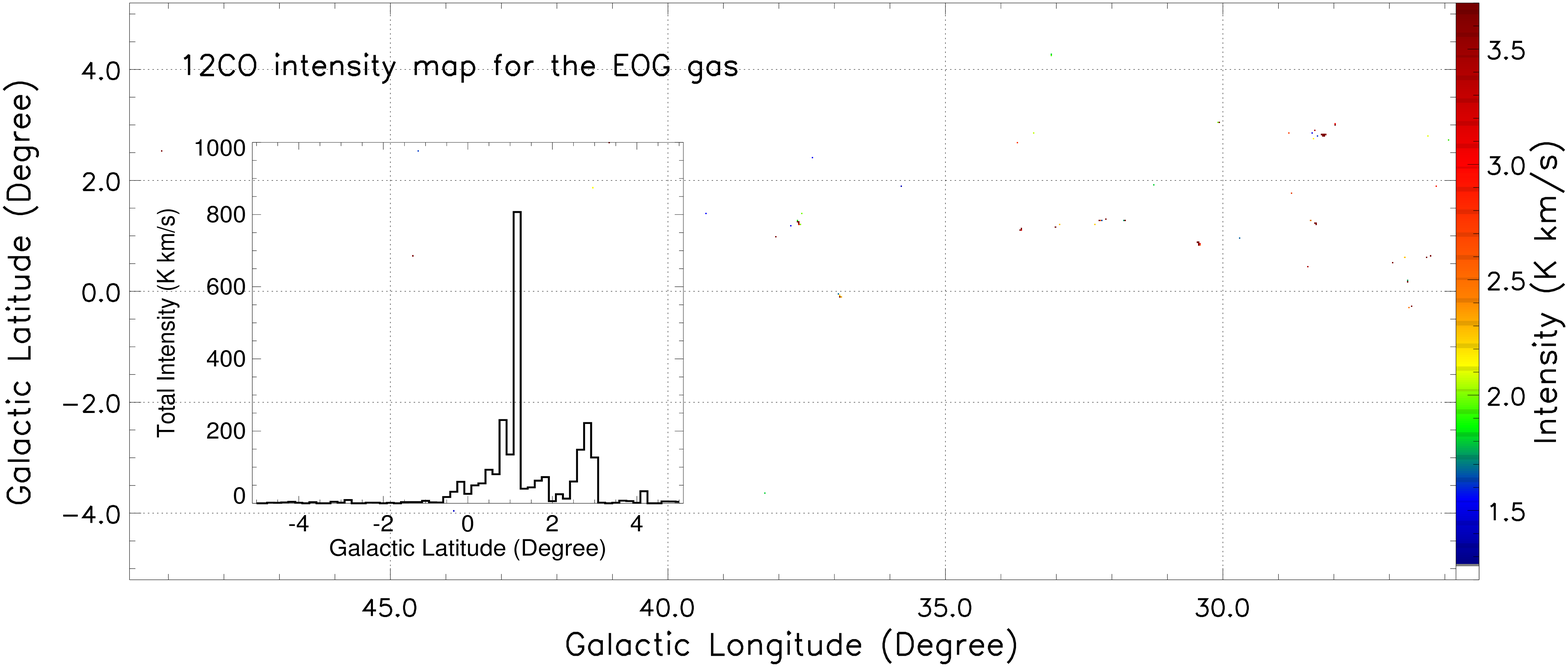}
\caption{
Distribution of the EOG \twCO\ emission with 
$V_{\rm LSR} \lsim-1.57 \times l+4.34$~km~s$^{-1}$. The CO intensity 
less than 1.2~K~$\km\ps$ (or $\sim 3\sigma$ for a typical velocity range of 4~$\km\ps$) 
is not shown on the map. 
Two intensity peaks are found to be at $b\sim$1\fdg2 and $b\sim$2\fdg8.
Note that the MC at \citep[$l=$40\fdg875, $b=$1\fdg250, $V_{\rm LSR} \sim -57\km\ps$; 
e.g., Table~1 in][]{2011ApJ...734L..24D} cannot be seen in the map
because of the velocity criterion here. But the MC probably belongs to the 
EOG. The corresponding CO emission of the distant MC can be seen in 
Figures~\ref{f2} and \ref{f9}.
\label{f10}}
\end{figure}

\begin{figure}[ht]
\includegraphics[trim=0mm 0mm 0mm 50mm,scale=0.8,angle=90]{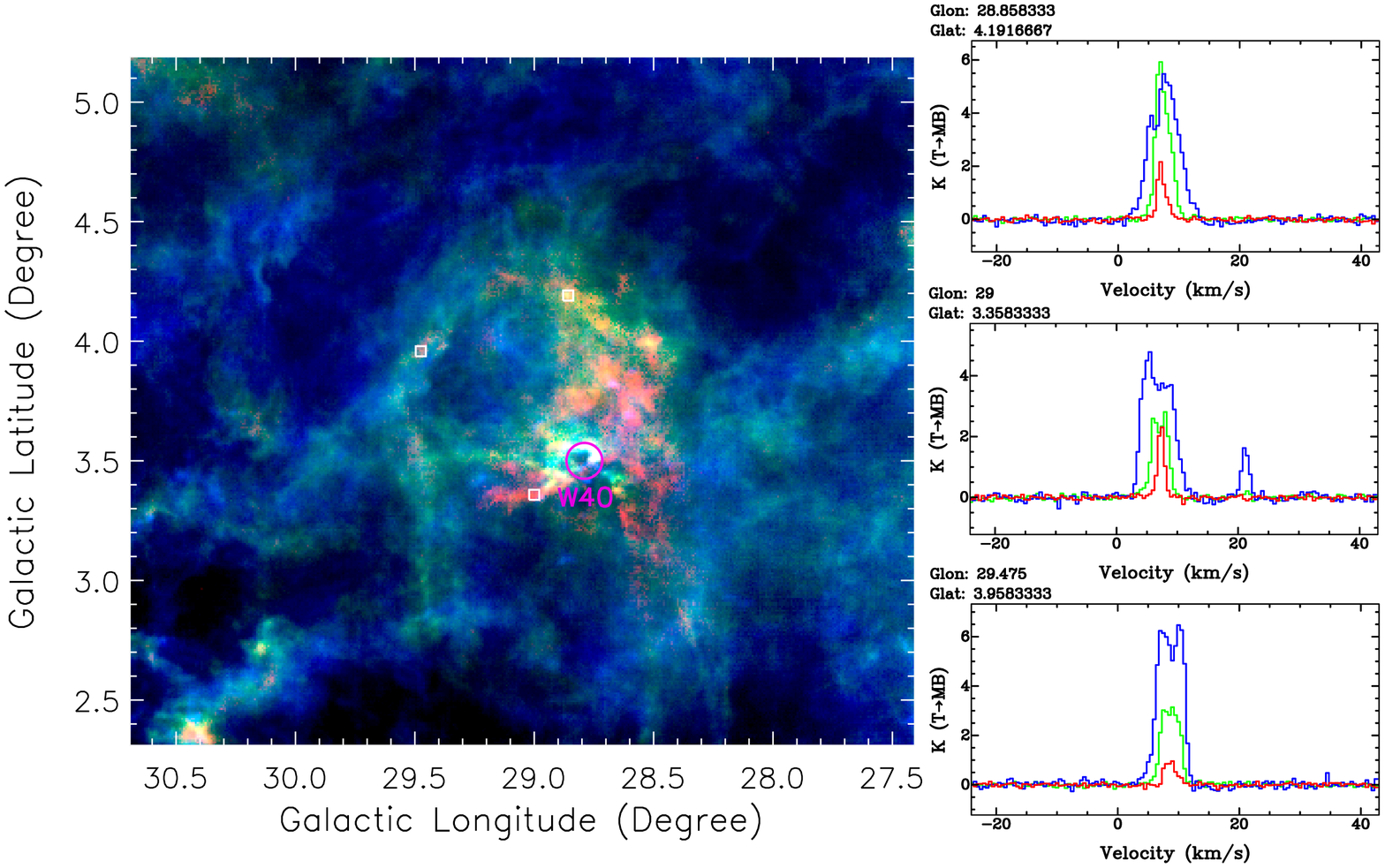}
\caption{
\twCO\ ($J$=1--0, blue), \thCO\ ($J$=1--0, green), and C$^{18}$O ($J$=1--0, red) 
intensity map in the 1--14~km~s$^{-1}$ interval toward the \HII\ region W40.
The \HII\ region W40 is marked as a purple circle centered at 
$l=$28\fdg8, $b=$3\fdg5 with a diameter of $9'$ \citep[e.g.,][]{2006ApJ...653.1226Q}.
Typical spectra, which are extracted from the three white boxes, are shown to
the right of the map.
\label{f11}}
\end{figure}

\begin{figure}
\gridline{
          \hspace{-10ex}
	  \fig{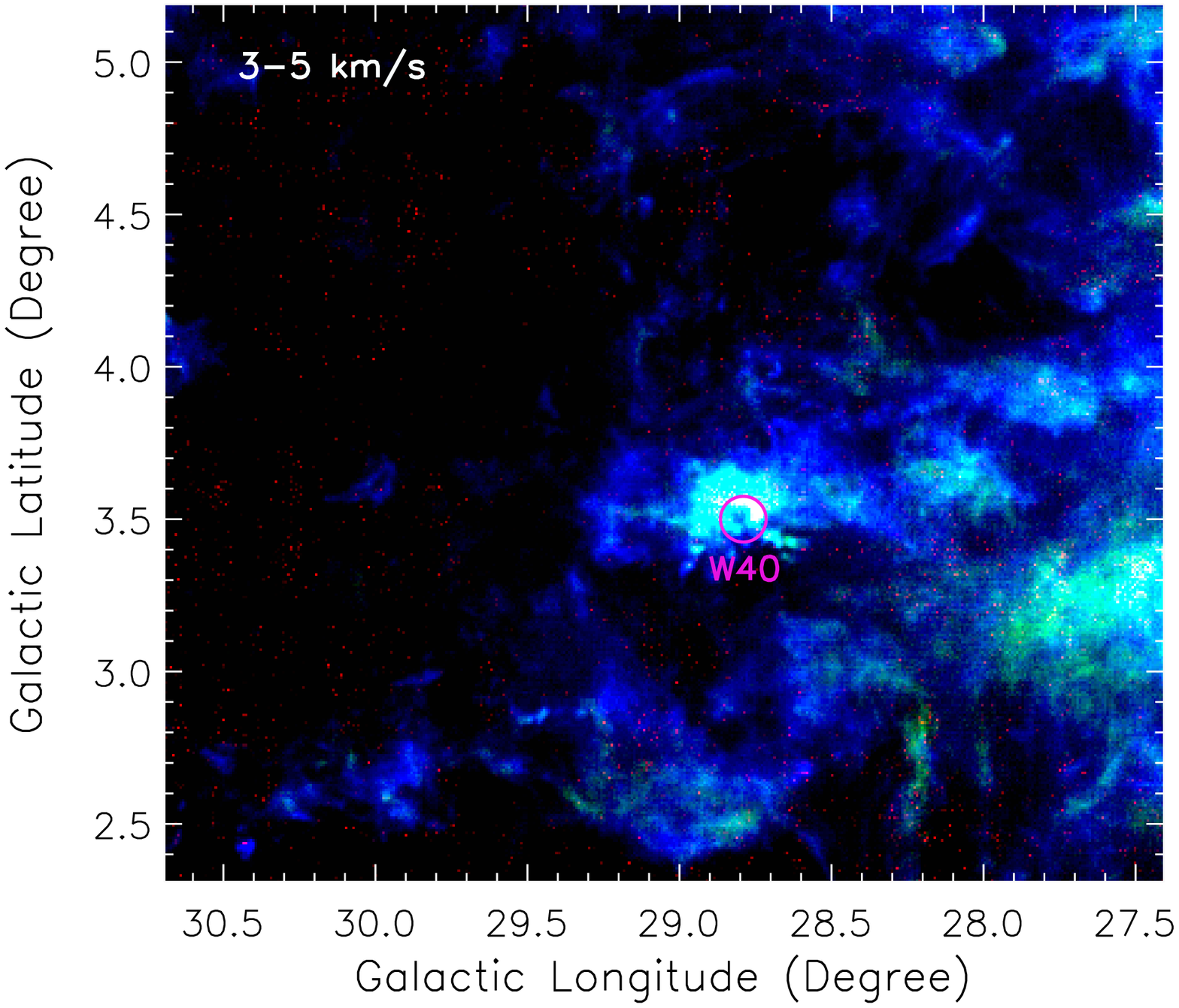}{0.6\textwidth}{}
          \fig{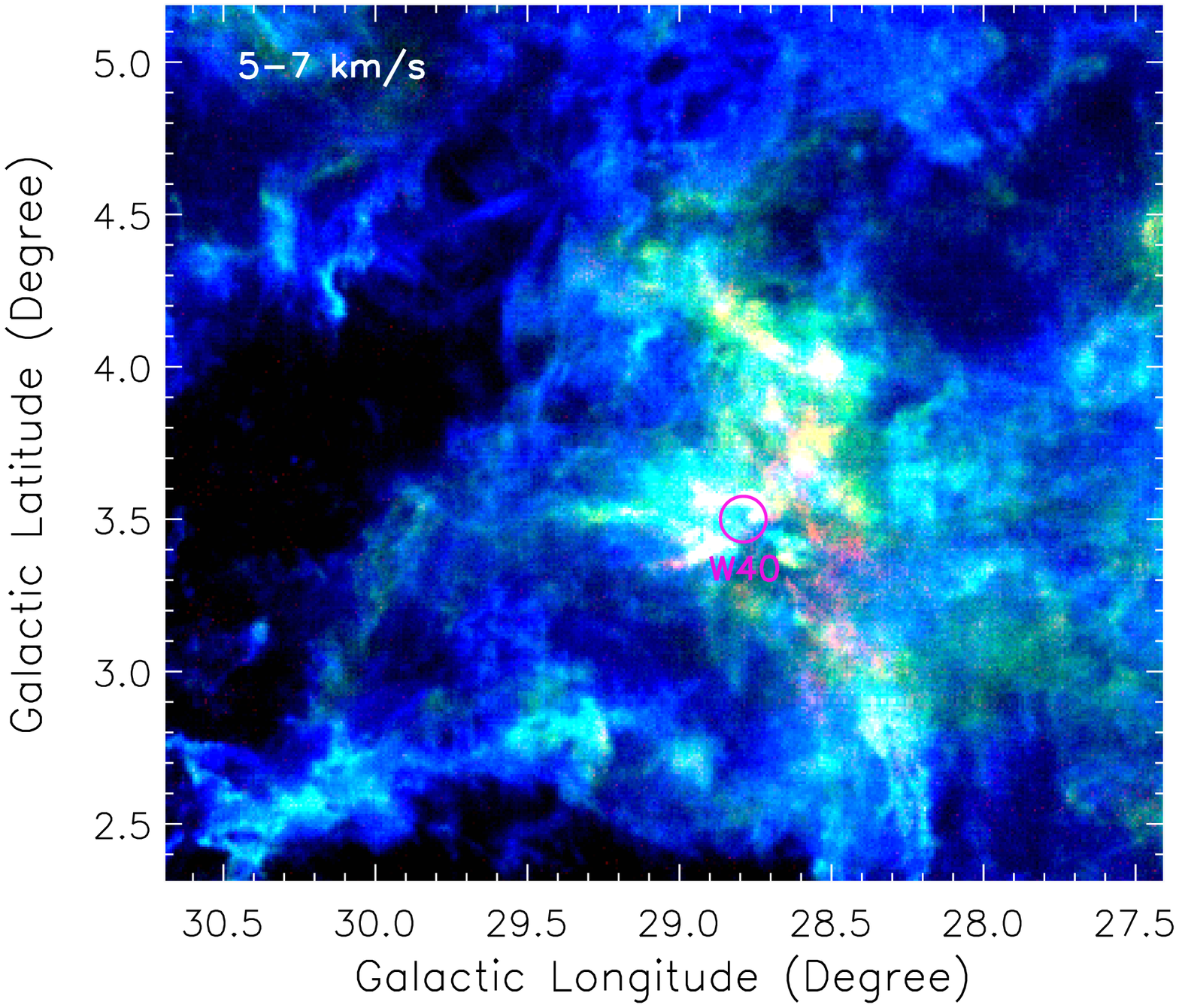}{0.6\textwidth}{}
          }
\vspace{-13ex}
\gridline{
          \hspace{-10ex}
	  \fig{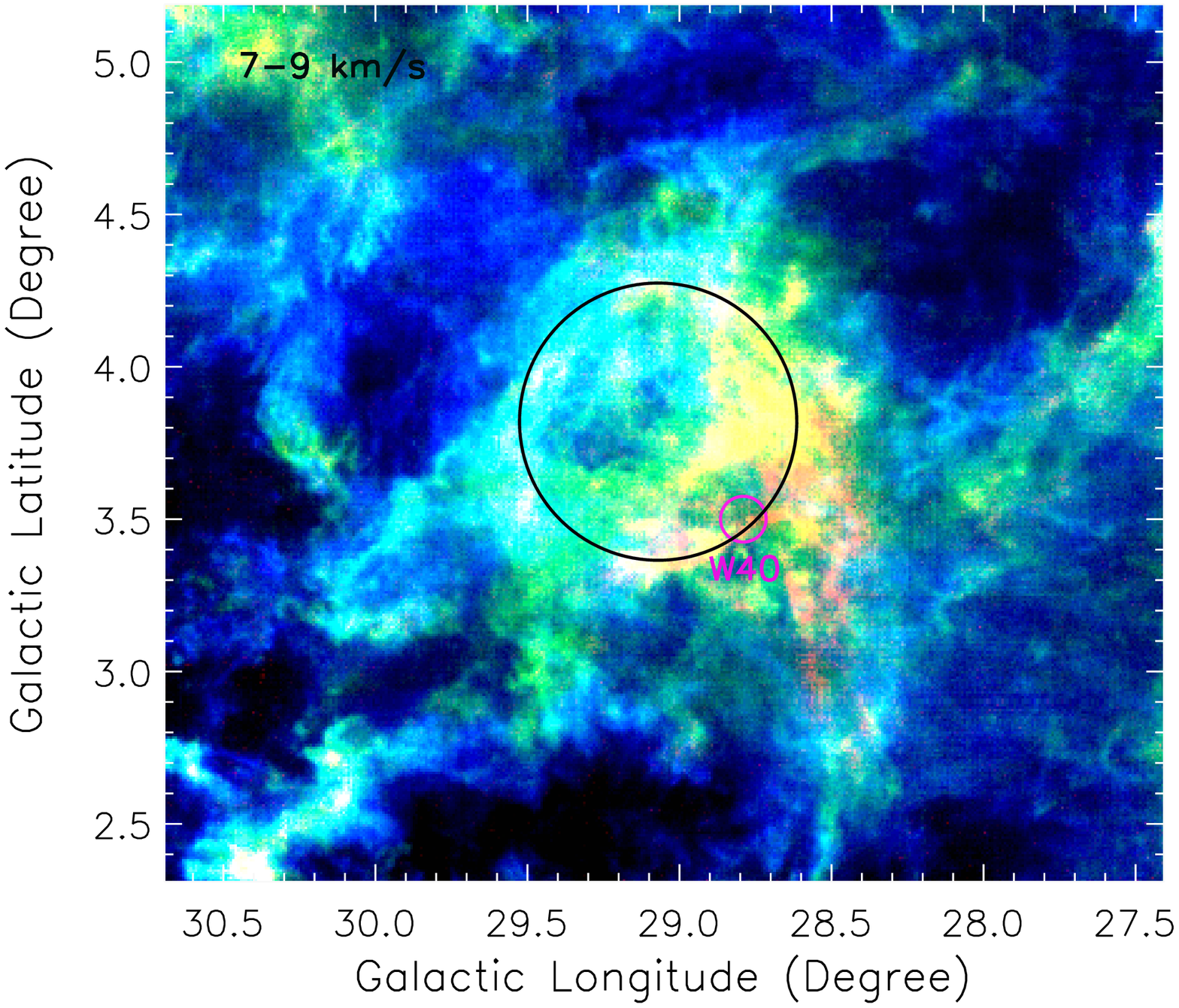}{0.6\textwidth}{}
          \fig{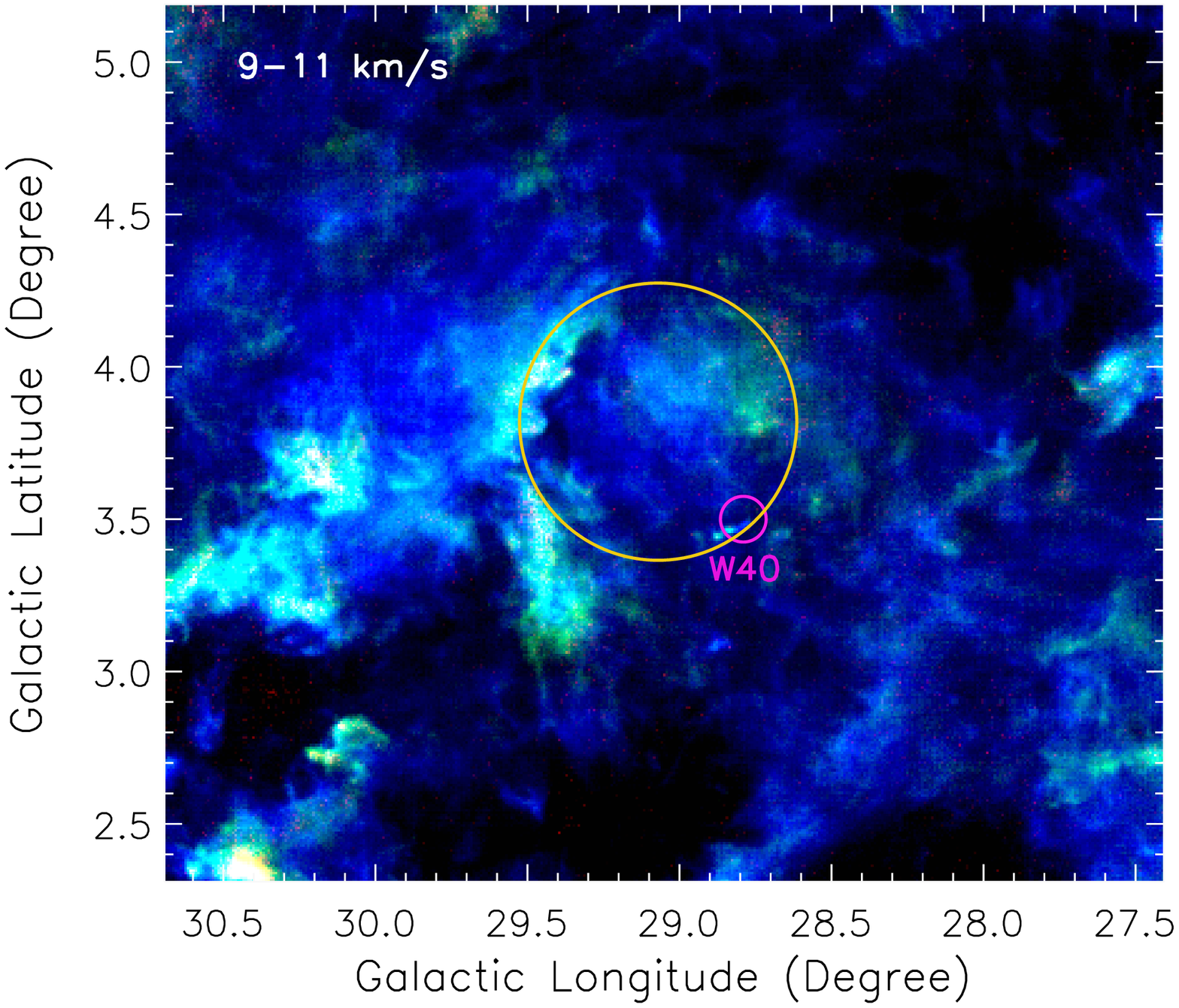}{0.6\textwidth}{}
          }
\caption{
Channel maps of \twCO\ ($J$=1--0, blue), \thCO\ ($J$=1--0, green), 
and C$^{18}$O ($J$=1--0, red) with velocity intervals of 3--5, 5--7,
7--9, and 9--11~km~s$^{-1}$
toward the W40 complex. The large circle, which is centered at 
($l=$29\fdg07, $b=$3\fdg82) with a radius of 0\fdg45, 
displays the bubble-like dense gas structure toward the complex.
\label{f12}}
\end{figure}

\begin{figure}[ht]
\includegraphics[trim=0mm 0mm 0mm 0mm,scale=0.5,angle=0]{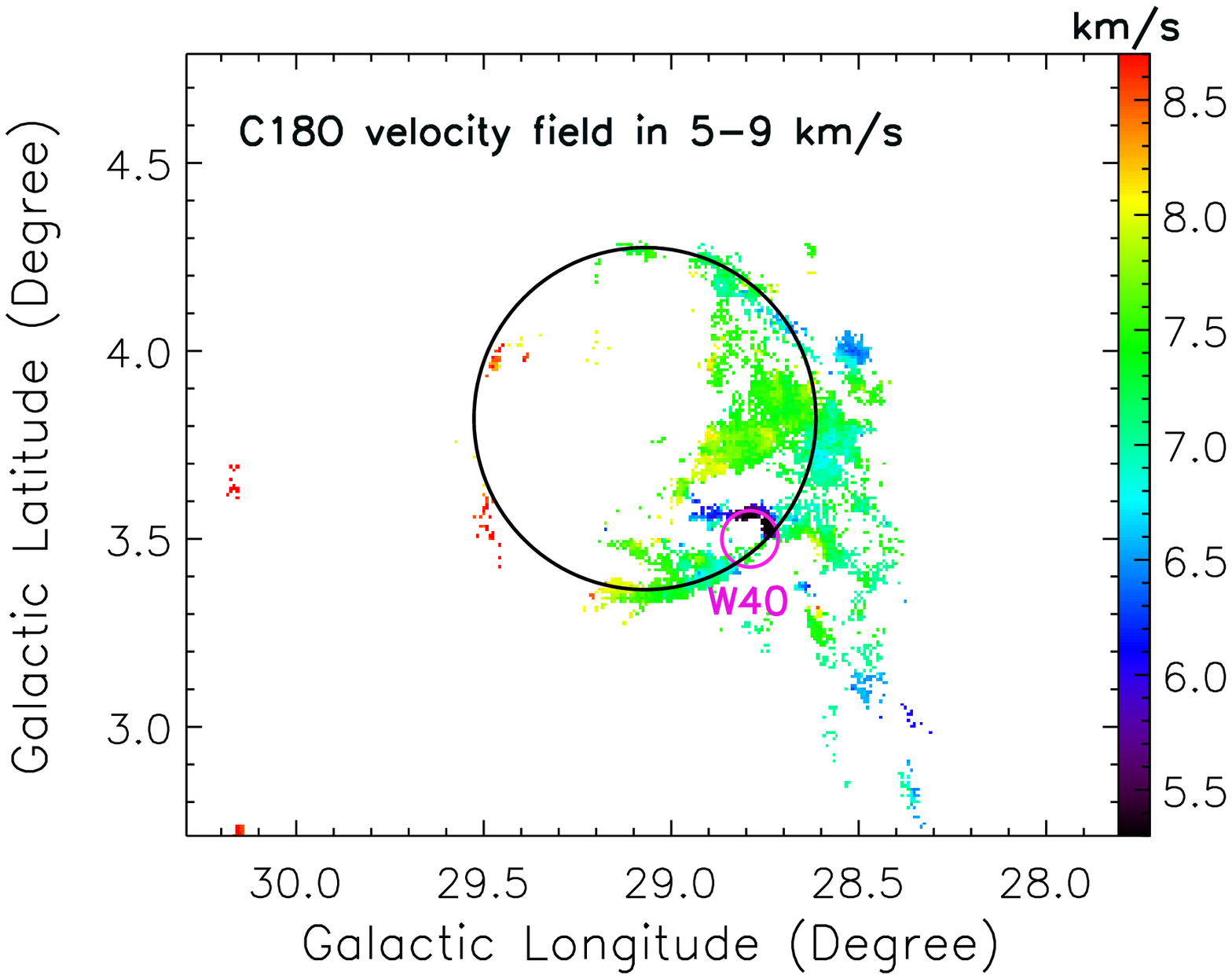}
\caption{
C$^{18}$O velocity field toward the W40 region. The big and small
circles are the same as those in Figure~\ref{f12}.
Samples with $T_{\rm MB} ({\rm C^{18}O}) \gsim 5\sigma$ are considered 
here.
\label{f13}}
\end{figure}

\begin{figure}[ht]
\includegraphics[trim=20mm 0mm 0mm 0mm,scale=0.45,angle=0]{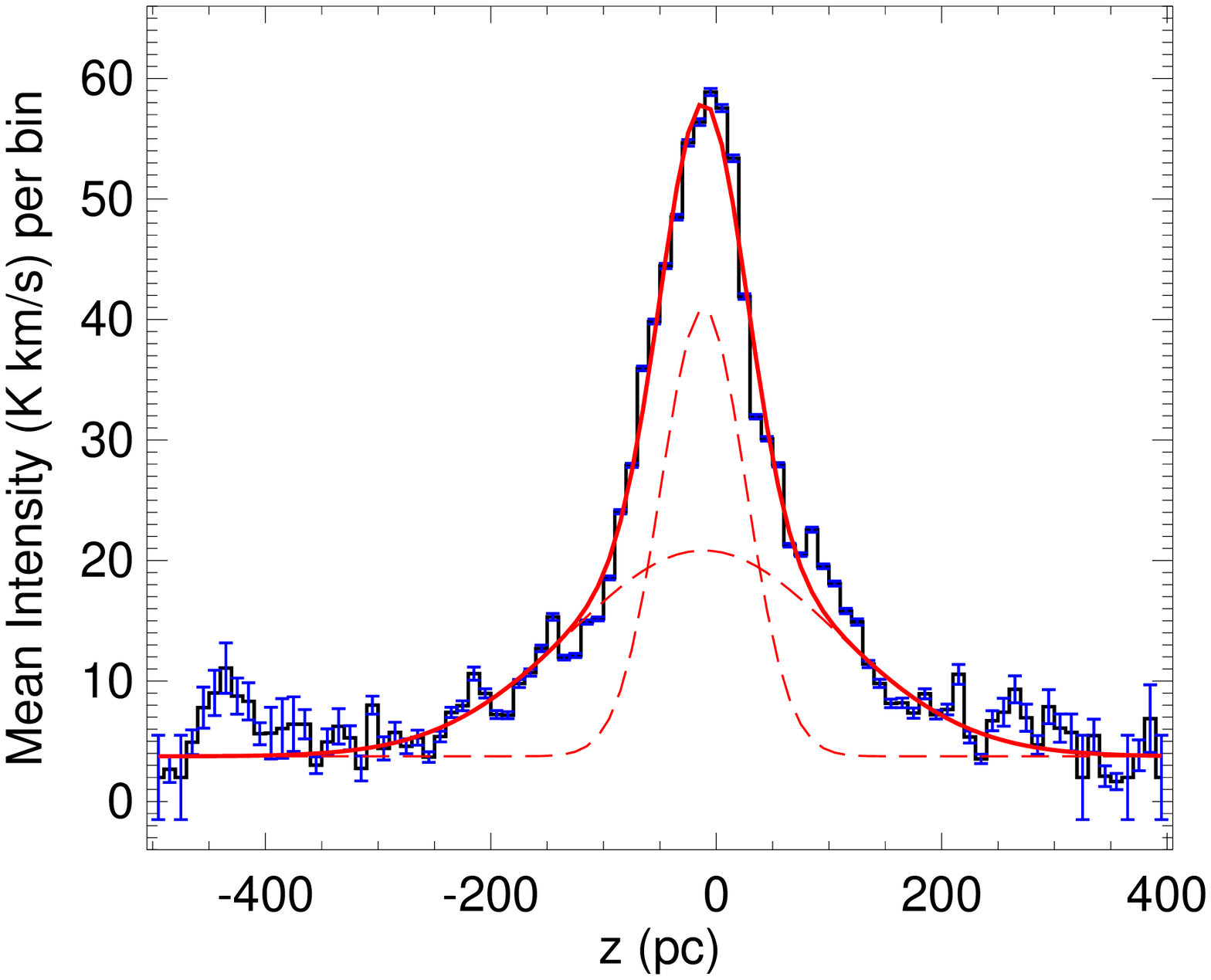}
\caption{
CO distribution along the distance from the Galactic plane of $b= 0^{\circ}$.
The two red dashed lines indicate the narrow and broad Gaussian components, respectively.
The error bars, which were calculated from 
$I_{\rm mean}$(CO)/(number of pixels per bin)$^{0.5}$,
are shown in blue.
Note that the zero-point of the best fitting is at 3.8~K~$\km\ps$
according to the MWISP CO data (see Table~2).
The unusual peak at $z \sim -435$~pc probably has a connection with
the dynamical interactions between the jets of the unique 
microquasar SS~433 in W50 and the surrounding high-$z$ gas (see the text).
\label{f14}}
\end{figure}

\end{document}